\documentclass[
 aip,
 amsmath,
 amssymb,
 reprint,
]{revtex4-1}

\usepackage{graphicx}
\usepackage{dcolumn}
\usepackage{bm}

\usepackage[utf8]{inputenc}
\usepackage[T1]{fontenc}
\usepackage{mathptmx}
\usepackage{etoolbox}

\usepackage{enumitem}
\usepackage{caption}
\usepackage{color}
\usepackage{xcolor}
\usepackage{soul}
\usepackage{verbatim}

\usepackage{xurl}
\usepackage{hyperref}
\hypersetup{
    colorlinks=true,
    urlcolor=blue
}

\usepackage{float}

\usepackage{graphicx}

\usepackage{array}
\usepackage{adjustbox}

\usepackage{tikz}
\usetikzlibrary{calc,shapes,arrows}

\usepackage[most]{tcolorbox}
\definecolor{myblue}{rgb}{0.31,0.44,0.75}

\makeatletter
\def\@email#1#2{%
 \endgroup
 \patchcmd{\titleblock@produce}
  {\frontmatter@RRAPformat}
  {\frontmatter@RRAPformat{\produce@RRAP{*#1\href{mailto:#2}{#2}}}\frontmatter@RRAPformat}
  {}{}
}

\makeatother

\begin{document}

\preprint{AIP/123-QED}

\title[ODIN: Open Data In Neurophysiology]{Open Data In Neurophysiology:\\Advancements, Solutions \& Challenges}

\author{Colleen J. Gillon$^{\dagger}$}
\affiliation{Department of Bioengineering, Imperial College London, London, SW7 2AZ, UK.}
\altaffiliation{$^{\dagger}$These authors contributed equally to this paper.}

\author{Cody Baker$^{\dagger}$}
\affiliation{Dartmouth College, Hanover, NH 03755, USA.}
\altaffiliation{$^{\dagger}$These authors contributed equally to this paper.}

\author{Ryan Ly$^{\dagger}$}
\affiliation{Scientific Data Division, Lawrence Berkeley National Laboratory, Berkeley, CA 94720, USA.}
\altaffiliation{$^{\dagger}$These authors contributed equally to this paper.}

\author{Edoardo Balzani}
\affiliation{Center for Computational Neuroscience, Flatiron Institute, New York, NY 10010, USA.}

\author{Bingni W. Brunton}
\affiliation{Department of Biology, University of Washington, Seattle, WA 98195, USA.}

\author{Manuel Schottdorf}
\affiliation{Princeton Neuroscience Institute, Princeton University, Princeton, NJ 08540, USA.}

\author{Satrajit Ghosh}
\affiliation{McGovern Institute for Brain Research, MIT, Cambridge, MA 02139, USA.}

\author{Nima Dehghani$^{*}$}
\affiliation{McGovern Institute for Brain Research, MIT, Cambridge, MA 02139, USA.}
\email{nima.dehghani@mit.edu}
\thanks{$^{*}$Corresponding author.}

\date{\today}

\begin{abstract}
Ongoing efforts over the last 50 years have made data and methods more reproducible and transparent across the life sciences. This openness has led to transformative insights and vastly accelerated scientific progress \citep{Grazulis2012, Munafo2017}. For example, structural biology \citep{Bruno2014} and genomics \citep{Porter2018, Benson2013} have undertaken systematic collection and publication of protein sequences and structures over the past half century. These data, in turn, have led to scientific breakthroughs that were unthinkable when data collection first began (e.g.~ \cite{Jumper2021}). We believe that neuroscience is poised to follow the same path, and that principles of open data and open science will transform our understanding of the nervous system in ways that are impossible to predict at the moment.\\ \\
New social structures supporting an active and open scientific community are essential \citep{Saunders2022} to facilitate and expand the still limited adoption of open science practices in our field \citep{Schottdorf2024}. Unified by shared values of openness, we set out to organize a symposium for Open Data in Neurophysiology (ODIN) to strengthen our community and facilitate transformative open neuroscience research at large. In this report, we synthesize insights from this first ODIN event. We also lay out plans for how to grow this movement, document emerging conversations, and propose a path toward a better and more transparent science of tomorrow.\\ \\
\textbf{Significance statement:} The adoption of open science practices is key to advancing research in neurophysiology. Drawing on insights from the inaugural Open Data in Neurophysiology (ODIN) symposium, this paper provides a much-needed discussion on the current state of open science in the field, identifying current needs of the community and providing a roadmap. Detailing both obstacles and solutions to the wide adoption of open science practices, it constitutes an important resource for the neurophysiology community on the path to open science.
\end{abstract}

\maketitle

\setcounter{footnote}{0}
\renewcommand{\thefootnote}{\fnsymbol{footnote}}

\small
\tableofcontents
\normalsize
\newpage 

\section{\label{sec:intro}Open Data in Neuroscience}
Over the past half-century, many subfields of the life sciences have undergone a profound transformation through the adoption of open data and open science practices. These shifts have catalyzed scientific breakthroughs in fields such as structural biology and genomics, with systematic data sharing accelerating discoveries that were previously unimaginable. The availability of large-scale protein structure databases has led to advances in protein folding prediction \citep{Jumper2021}, while the widespread adoption of genomic repositories has revolutionized our understanding of genetic variation and disease \citep{1000Genomes2015}. In constrast, neuroscience, and neurophysiology in particular, is still in the early stages of adopting open science values and practices.

Despite the vast amounts of neurophysiological data being generated, significant challenges remain in standardizing, sharing, and integrating these datasets across research groups. The potential benefits of open neuroscience are clear: improved reproducibility, broader collaboration, efficient data reuse, and deeper insights into the fundamental workings of the brain. Yet, barriers such as data heterogeneity, resource limitations, and institutional inertia continue to slow progress. Recognizing this gap, we organized the Open Data in Neurophysiology (ODIN) symposium to bring together researchers, data scientists, and policymakers to discuss strategies for advancing open data practices in neurophysiology\footnote{For the full video versions of ODIN 2023 session recordings, see \url{https://www.youtube.com/playlist?list=PLQVnU1OJzOn_mFlUL8aWQym4HfVvhlrGE}.}.

\begin{figure*}[htbp]
  \centering
  \includegraphics[width=0.95\linewidth]{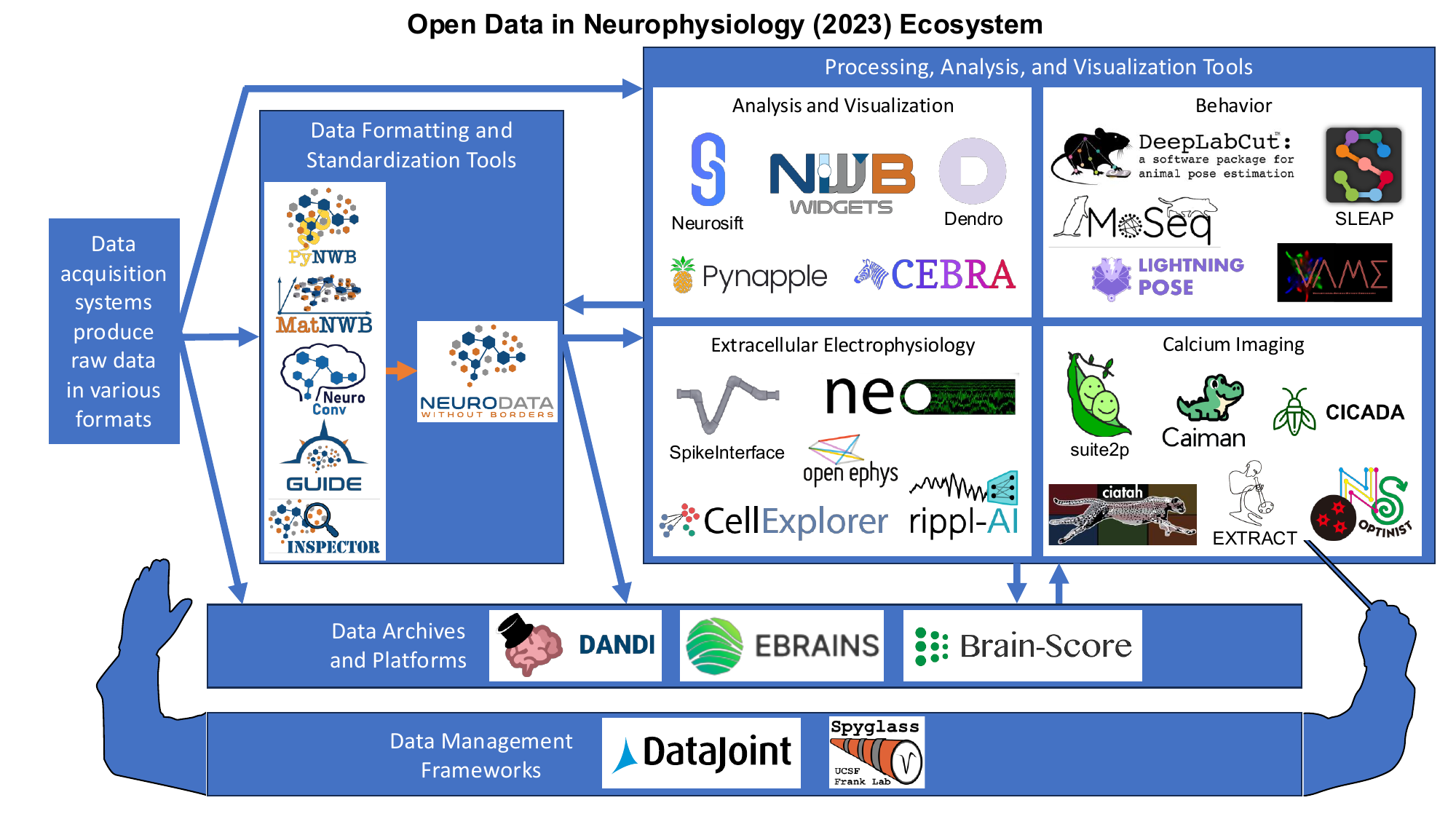}
  \caption{Snapshot of the current ecosystem of open source neurophysiology tools. Note that the ecosystem is growing, and thus larger than what is depicted here. Included here are example tools that were presented or discussed during the ODIN 2023 symposium. See Table~\ref{tab:toolkits} for more information about each toolkit.}
  \label{fig:ecosystem}
\end{figure*}

This paper synthesizes and expands upon key insights from the symposium. First, we explore the vast ecosystem of existing neurophysiology tools and resources (Fig.~\ref{fig:ecosystem}). Specifically, we discuss devices, neuroinformatics and platforms, followed by knowledge extraction, software and modeling in neurophysiology, in each case presenting recent advances, as well as some of the major obstacles to progress in these domains. Next, we dive more deeply into current technical challenges in neuroinformatics that may be hindering widespread adoption of open science practices, also discussing potential solutions. Lastly, we provide a forward looking perspective highlighting the current needs of the open science community, as well as recommendations for individuals and organizations seeking to engage with and promote open science practices. Throughout each of these sections, certain central themes consistently reemerge: (1) the need for robust infrastructure and standardized formats to enable seamless data sharing, (2) the cultural and institutional shifts required to foster openness and collaboration, and (3) the role of computational tools and machine learning in making large-scale neurophysiological data more accessible and interpretable. By identifying actionable steps and promising ongoing initiatives, we aim to provide a roadmap for researchers, institutions and fundung agencies seeking to contribute to a more transparent and collaborative scientific ecosystem.

\section{\label{sec1}Devices, Neuroinformatics, and Platforms}
\subsection{\label{sec1:ses1}New Devices and High Throughput Acquisitions}

Recent developments in neurotechnology have significantly improved the spatiotemporal resolution and throughput at which brain activity can be recorded. In turn, however, this growth in available data has created significant challenges for data management, sharing and long-term storage.

\subsubsection*{Advances \& Examples}

Advances in devices and acquisition systems are often the result of significant technical and engineering breakthroughs. For example, the development of multi-thousand channel electrocorticography (ECoG) grids now enables much denser and higher resolution mapping of brain activity than was possible using traditional clinical electrodes. This innovation was made possible by the development of smaller electrodes which can still yield a high signal-to-noise ratio. It was also critically facilitated by technical advances in thin-film microfabrication which allows electrodes to be more stably arranged along the brain's curvilinear surface. The move towards wireless systems has also increased the efficiency of acute and chronic monitoring, while also reducing its intrusiveness \citep{Tchoe2022}. Similarity, the development of Neuropixels, a silicon probe which allows high-density simultaneous recording the activity of hundreds of neurons in awake and freely moving animals, has revolutionized the field of systems neuroscience, in which this technology is now ubiquitous \citep{Jun2017}. The latest version of the probe, Neuropixels NXT, aims to be more compact in design to increase the detail and scale at which electrophysiological neural activity can be recorded. 

In optical imaging, the use of light sculpting and temporal multiplexing now enables cortex-wide, volumetric recording of neuronal activity across millions of neurons at single-cell resolution. Data yielded by this type of imaging technique will be critical to studying patterns of coordinated activity across distant regions of the brain \citep{Demas2021}. On a more fine-grained scale, combining voltage-sensitive fluorescent proteins activated by red light with blue-light-activated channelrhodopsins for neuronal stimulation now enables optical monitoring of electrical activity within neurons at high spatial and temporal resolution \citep{Adam2019}. Since this technique allows spikes and subthreshold voltages to be resolved, the data it yields could greatly enhance our understanding of within-neuron voltage dynamics, input-output relationships in local networks, and the plasticity rules that govern them. 

Advances in tools like these facilitate the collection of vast amounts of high quality neural data, and have the potential to yield transformative insights into the dynamics that govern the activity of entire neuronal populations.

\subsubsection*{Challenges \& Concerns}

Arising directly from this significant improvement in resolution and throughput, however, is the challenge of managing very large datasets. With researchers increasingly collecting datasets comprising terabytes (TBs) of raw data, reliable large-scale data management systems are needed to store and disseminate them. If a researcher or laboratory opts to preserve all raw data collected, they must either devise their own data management system or rely on shared or externally maintained ones, such as institutional data management systems or large-scale online repositories like the DANDI Archive. Online data repositories in particular offer a very attractive solution. In addition to reducing the pressure for individual researchers to manage long-term data storage themselves, they are ideal for data dissemination. Of course, the challenge of data management remains, being simply displaced onto the managers of these online repositories. Comparisons with data management practices at institutions such as CERN—which handles 50–100 petabytes annually \citep{CERN2025}—suggest that current neurophysiological data repositories should be able to cope for now. However, if, over the next decades, the number of researchers around the globe opting to share their datasets publicly increases substantially, as we hope it will, these repositories will need substantial and reliable resources to scale at a rate that matches demand.

In addition to ensuring that repositories have the resources needed to scale their data management capacities, it is also important to consider carefully, as a community, what types of data should be preserved and what types of data can safely be discarded. Indeed, an alternative to storing TBs of raw data is to keep only a pre-processed, and thus much smaller, version of a dataset. Of course, this is not always appropriate, but there may be many cases where this approach provides an adequate compromise between data reusability and responsible use of resources. In such cases, a critical challenge for researchers is ensuring that the adopted compression and pre-processing steps do not discard information critical for future validation, reproduction of results or reanalysis of the data. For example, spike sorting methods must be carefully tuned to avoid yielding too many false positives or false negatives, complicating the extraction of reliable spike trains \citep{Zhang2023}. Thus, although pre-processed spiking data are much easier to share and reuse, critical information can easily be lost if the raw dataset they were extracted from is not also preserved. Guidelines for when and how to share pre-processed rather than raw datasets, or even for when to share protocols and analysis pipelines instead of datasets, are needed to ensure that storage resources are used responsibly, storage needs are met over the long term, and the vital role of publicly available data in reanalysis and future discovery is protected \citep{Dehghani2024}.

A related concern is the trade-off between scale and interpretability. While recording from large populations of neurons can greatly increase the information contained in a dataset, this may not automatically lead to better scientific insights. As dataset sizes grow, so does the complexity of the data and often of the associated analyses. New analytical frameworks that move beyond traditional single-neuron analysis are needed to extract reliable network-level interpretations of brain activity. Collaborative efforts in this direction between experimentalists, theorists, data scientists, and engineers will be needed to ensure that neuroscience remains both data-rich and scientifically insightful. 

In summary, high throughput neurophysiology offers unprecedented opportunities for understanding brain function. However, these advancements are accompanied by equally significant challenges. Only through a concerted effort that integrates advanced technologies with robust data management and scalable analytical strategies can the full potential of these innovations be realized.

\subsection{\label{sec1:ses2}Neuroinformatic Resources}
A major goal of open science is to promote the sharing of datasets, protocols and analysis tools across laboratories. Such practices reinforce transparency and scientific integrity, and can greatly increase research consistency and quality across laboratories. From an ethical point of view, measures that maximize data reuse also reduce the need for animal experimentation, thus directly contributing to animal welfare. However, a major impediment to resource sharing is the diversity of data formats, workflows and coding practices used by different groups. Common standards and guidelines are needed to manage the ever-growing volume and complexity of neurophysiology data discussed above. In response, the landscape of neuroinformatics tools is undergoing substantial transformations, driving an increased adoption of common data standards, the growth of standardized data repositories, and the development of interoperable workflow and data visualization tools.

\subsubsection*{Advances \& Examples}

Ensuring that data collected and shared by one laboratory can easily be reused by a different group requires adopting and investing in common data standards. The Neurodata Without Borders (NWB) ecosystem has emerged as a robust, multidisciplinary framework for organizing diverse datatypes—from neural activity recordings to experimental metadata—into a single, hierarchical format. Recent developments include support for cloud-based data access, as well as the integration of external resources and of a variety of application programming interfaces (APIs). These aim to increase usability and lower barriers to adoption for users from different backgrounds. In particular, more intuitive tools such as NeuroConv and NWB GUIDE should simplify the process of converting proprietary data into the NWB format. Once standardized, these datasets can be stored and shared on repositories such as DANDI (Distributed Archives for Neurophysiology Data Integration) which, as of mid 2025, housed nearly a petabyte of data \citep{dandisize2025}. DANDI has also launched an associated JupyterHub platform to encourage cloud-based data analysis, further supporting the reusability of the data it hosts. This expansion of data standards and repositories hopefully signals a broad shift toward open science, where, in line with FAIR principles, data is \emph{Findable, Accessible, Interoperable and Reusable} \citep{Wilkinson2016}.

Complementing these advances in data standards and repositories are innovations in computational workflows and web-based visualization. With platforms like DataJoint, laboratories can recruit the assistance of external experts to create end-to-end computational workflows covering the entire lifecycle of their neuroscience projects—from data acquisition and animal management to spike sorting and behavioral analysis \citep{Yatsenko2018, Yatsenko2021}. DataJoint resources can also be used to evaluate the operational maturity of existing workflows. Open-source solutions can then be identified to improve these workflows, for example by automating time-consuming steps using reliable AI-based tools. Web-based tools are also changing how data can be analyzed and visualized. Open-source platforms such as Figurl\footnote{For more information on Figurl, see \url{https://github.com/flatironinstitute/figurl}.}, Neurosift \citep{Magland2024}, and Dendro\footnote{For more information on Dendro, see \url{https://dendro.vercel.app/}.} offer interactive, shareable visualizations that integrate seamlessly with data standards like NWB and repositories like DANDI which use these standards. These tools not only simplify data exploration, but also enhance collaboration as they enable scientists to share interactive figures and analyses through URLs. Altogether, automated workflows and containerization technologies show great promise in enhancing reliability, reproducibility and transparency in scientific research.

\subsubsection*{Challenges \& Concerns}

Amid these promising developments, several challenges still remain. One challenge is ensuring that new technologies and resources can be easily adopted by different laboratories operating in very diverse research environments. Researchers establishing new laboratories should be encouraged to consult early with resource developers and incorporate standardized workflows from the beginning to streamline data management and analysis. For existing laboratories with well-established procedures, overcoming barriers to adoption will likely be much more difficult. In both cases, enhancing usability and providing robust user support remains critical. Importantly, standardizing commercially available data acquisition systems has the potential to greatly facilitate data sharing and analysis, removing most of the onus of data standardization from individual researchers. Notably, unifying these systems also presents potential advantages for clinical settings as it could streamline the management of data identifiability and privacy, thus reducing the risk of error and mismanagement.

By integrating standards like NWB, repositories such as DANDI, and innovative computational and visualization tools, the neuroinformatics community is making significant strides toward addressing the technical challenges of modern neuroscience research. Collaborative efforts between experimentalists, data scientists, and engineers are required to ensure that these new and developing tools can be productively adopted by the community.

\subsection{\label{sec1:ses3}Platforms \& Collaborative Initiatives}

It is also increasingly recognized that improvements in collaborative platforms and research infrastructures are critical to addressing the reproducibility crisis and advancing our understanding of neurophysiology. Shared tools like integrated databases and common analytical frameworks are invaluable, openly accessible resources for the global scientific community. They can also bring together more distant research subfields, enabling, for example, cross-species neurophysiology comparisons. The current move towards inclusive, transparent, and collaborative research infrastructures is already accelerating discoveries and driving innovation in neuroscience.

\subsubsection*{Advances \& Examples}
The Japan Marmoset Initiative, the OpenScope platform, the Allen Institute for Neural Dynamics, and the International Brain Laboratory exemplify a shift toward large-scale, collaborative neuroscientific projects. The Japan Marmoset Initiative is a multi-institution initiative using data from genetically modified marmosets to advance brain mapping and disease modeling research. This initiative has yielded a detailed database of 3D structural, diffusion, and resting-state magnetic resonance imaging (MRI) data, as well as an \textit{in situ} hybridization-based gene atlas. Projects using this data have the potential to yield significant insights, facilitating interspecies comparisons and enhancing our understanding of disease mechanisms such as those underlying Rett syndrome \citep{Hata2023,Okano2015,Skibbe2023}.

A new platform inspired by astronomical observatories has also recently emerged. With the OpenScope platform\footnote{For more information on the OpenScope platform, see \url{https://alleninstitute.org/division/neural-dynamics/openscope/}.}, the Allen Brain Observatory has made their high throughput pipelines for two-photon microscopy and Neuropixels recordings available for public use in order to democratize access to cutting-edge data collection. Researchers worldwide whose proposals are accepted receive neurophysiology data collected specifically for their project under highly standardized conditions. The double-blind review process for proposals, inspired by approaches to allocating time on shared telescopes, aims to ensure that resources are allocated based on scientific merit and equity considerations to foster both innovation and inclusivity \citep{DeVries2023,Koch2022}. Complementary tools like the OpenScope Databook have also been developed to help standardize and democratize data analysis and visualization techniques.\footnote{For more information on the OpenScope Databook, see \url{https://alleninstitute.github.io/openscope_databook/intro.html}.} Efforts at the Allen Institute for Neural Dynamics have focused on both data management and computing for large-scale neuroscience.  On the data management side, the institute aims to make data compliant with FAIR guidelines at acquisition, through robust metadata capture and early conversion to common data standards such as BIDS, NWB, and OME \citep{Goldberg2005, Gorgolewski2016}. They also aim to harness cloud computing infrastructures to reduce logistical hurdles related to moving and storing data, and to enable complete software and hardware environments to be shared via containerized pipelines. As these resources are publicly shared, these efforts will accelerate access to high quality, standardized data and to fully reproducible workflows that can operate both in the cloud and locally.

Notably, these advances are directly aligned with the aims of the United States' National Institute of Health (NIH)'s BRAIN (Brain Research through Advancing Innovative Neurotechnologies) Initiative, as outlined in the BRAIN 2025 report. Indeed, a core aim was to create of a public, integrated ecosystem for seamless sharing of datasets and analysis tools. This ecosystem is currently distributed across a broad range of platforms which, collectively, house thousands of datasets (see Tables \ref{tab:brain_archives} and \ref{tab:archives}). These platforms are poised to grow even further with the gradual introduction of mandates, like the BRAIN Data Sharing Policy, requiring researchers to share their data and code alongside their publications\footnote{For more information on BRAIN 2025, see \url{https://braininitiative.nih.gov/vision/nih-brain-initiative-reports/brain-2025-scientific-vision}.}. Together, the measures advances by the BRAIN Initiative aim to ensure that research tools are shared broadly and that research data are made available to the wider community in a timely manner.

The International Brain Laboratory (IBL) has built a brain-wide map of neuronal activity during behavior by deploying standardized experimental protocols across 22 collaborating laboratories\footnote{For more information on the IBL, see \url{https://www.internationalbrainlab.com/}.}. This initiative has generated an extensive dataset from nearly 33,000 neurons recorded using Neuropixels probes, offering an unprecedented opportunity to rigorously investigate the distributed processing of sensation, decision-making, action, and prior beliefs in a single task \citep{IBL2021,IBL2023,Findling2023}. Given the standardized, but distributed approach of the IBL, this dataset also provides an opportunity to better understand key issues of reproducibility in neurophysiology. Notably, analyses of the IBL dataset have shown that although standardized data collection can greatly reduce variability, even the use of different analytical methods and metrics on the same dataset can meaningfully alter the results of an experiment and their interpretation \citep{IBL2024}. Altogether, the IBL's approach demonstrates the value of developing unified and sound methodologies covering the full lifecycle of an experiment, from data collection to analysis, and of harnessing open science practices to address complex neurophysiological questions and tackle the challenge of reproducibility.

Together, initiatives like these have the potential to considerably advance neurophysiology research. Given their scale, they are generally able to harness cutting-edge technologies and tackle complex operational and analytical challenges much more easily than individual laboratories can. Their commitment to open and reproducible research has also meant that the comprehensive, integrated databases and resources they have created are openly accessible to the global neurophysiology community. This offers critical opportunities for individual researchers worldwide, who may not have the requisite resources, data collection tools or software engineering capabilities, to contribute their expertise to the creation and analysis of complex datasets, and thus can help foster more inclusion and collaboration across the scientific community.

\subsubsection*{Challenges \& Concerns}

The challenges faced by large-scale platforms overlap heavily with those faced by individual laboratories. However, given the specific mandate of several of these platforms to collect and disseminate vast datasets with high reuse potential for the broader research community, the consequences of not sufficiently addressing these challenges are arguably scaled up for these large platforms. Finding solutions that work for individual research groups, but can also scale to larger organizations, is therefore critical to ensuring that the deployment of large-scale platforms is effective, sustainable and constitutes an appropriate use of resources.

A key concern is again the issue of data management. As discussed in detail in the section on high throughput neuroscience, challenges include deciding what types of data, whether raw or processed, to preserve, determining how best to make datasets publicly available, and ensuring that the analytical frameworks being used are appropriate for the complexity and scale of the datasets. Another difficulty, also shared with individual laboratories, is the challenge of ensuring that the datasets collected by large-scale platforms include comprehensive metadata. Properly capturing, recording, and disseminating detailed metadata—including environmental conditions, experimental protocols, and subtle procedural nuances—is essential for contextualizing experimental outcomes and ensuring data can be appropriately reused, reproduced, and interpreted. Yet, tools for doing this easily and consistently are still lacking. Thus, for both large-scale and individual research efforts, there is a pressing need for user-friendly metadata collection and management tools. Lastly, an additional challenge, particularly faced by large-scale platforms handling human data, is ensuring that data privacy concerns, particularly in clinical settings, are properly addressed and that the deployment of open data practices does not compromise confidentiality. 

To effectively address these challenges, experimentalists, data analysts and software developers must collaborate on streamlining workflows in a way that is scalable, and preserves flexibility where necessary. Automated tools, when properly developed, can make the collection, processing and dissemination of data more standardized and robust, thereby reducing risks of data or metadata loss and better protecting data privacy. However, flexibility remains important as scientific needs can vary greatly between and even within experiments. Additionally, more centralized technical support platforms are needed at the intermediate level within institutions as resources for individual laboratories. Such core facilities can provide much needed expertise, assisting researchers in leveraging complex, emerging technologies without becoming overwhelmed by complex technical implementation details.

\section{\label{sec2}Knowledge Extraction, Software, Modeling}
\subsection{\label{sec2:ses1}Experimental Models and Data-Driven Approaches}
Innovations in experimental design and analysis techniques also have the potential to significantly accelerate discovery and generate new knowledge in the field. However, to achieve this potential, these advances and the data they yield must be effectively disseminated to the community. We extend the discussion here into examples of new experimental models, as well as the collection and use of cutting-edge neurophysiology datasets, discussing how, through interdisciplinary efforts, these could be used to address long-standing questions in neuroscience about neural circuit development, the processing of complex sensory information, and the brain’s ability to generalize compared to artificial intelligence systems.

\subsubsection*{Advances \& Examples}

A recent innovation that has the potential to greatly improve our ability to study brain-like neural circuits using in vitro techniques is the development of cortical organoids. These three-dimensional structures capture key characteristics of real brains, like their diverse cell types and synaptic properties, which dissociated cultures typically cannot. Using optofluidic-CMOS multielectrode arrays and Neuropixels recordings, researchers have demonstrated the emergence of neuronal assemblies with sufficiently structured oscillatory dynamics to generate local field potentials \citep{Sharf2022}. This technology has the potential to be of great benefit to the community, opening up new avenues for exploring intrinsic cortical activity, network dynamics, and complex processes like learning and memory formation under well-controlled and replicable conditions.

In the study of sensory processing, closed-loop data-driven model training in which data is collected at the same time as neural networks are trained has emerged as a promising new technique. An example is the use of ``inception loops'' where stimuli presented to mice are modified in real time using deep learning in order to identify the stimulus patterns that most strongly excite specific neurons \citep{Walker2019}. Since the preferred stimulus patterns predicted for individual neurons are fed back into the experiment, the quality of the network's predictions can be directly evaluated, thus avoiding some of the problems that arise when trying to interpret neuronal activity based on the outputs of black-box neural networks. Techniques like this are needed to move beyond overly simple interpretations of neuronal activity. 

Large, high-quality multidimensional datasets also provide key opportunities to study complex neural processes and develop data-driven models. AJILE12 (Annotated Joints in Long-term Electrocorticography for 12 human participants) is an electrocorticography dataset collected in humans engaging in video-recorded and annotated natural behavior. It is openly available in NWB format and can be explored using a browser-based dashboard. When available to the broader neuroscience community, data exploration pipelines like these drive progress in a wide range of fields. For example, AJILE12 can be used for basic and clinical research, deepening our understanding of the relationship between neural activity and motor behaviors on the one hand, and enabling us to develop more robust brain-computer interfaces on the other \citep{Peterson2021, Peterson2022}. Another highly reusable dataset is the MICrONS project dataset, described in previous sections, which provides a unique opportunity for researchers to study the link between structural connectivity and functional output in the visual system. Datasets like AJILE12 have also been integrated into educational initiatives like Neuromatch Academy \citep{vanViegen2021} providing researchers in training the opportunity to cut their teeth on datasets and tools that are actually used in their field\footnote{For more information on AJILE12, see \url{https://github.com/neurovium/Neuromatch-AJILE12}.}.  

In addition to large-scale datasets, smaller-scale datasets collected by individual laboratories are critical to advancing research. The highly diverse research focuses of different groups ensure that together we are collecting a wide array of puzzle pieces. For example, the role of individual cell types in shaping circuit function in sensory cortices remains poorly understood, and across laboratories, the focus and methods of investigation often differ considerably. Research combining two-photon imaging with optogenetic control over specific cell types has helped illuminate how different inhibitory neuron populations shape aspects of auditory processing, like frequency discrimination and adaptation to temporal sound patterns \citep{Natan2015, Tobin2025}. If the infrastructure needed for laboratories to easily share and explore each other's datasets and pipelines can be made available, the gaps between the puzzle pieces explored by different groups can be better bridged. This will pave the way for us as a community to obtain a much fuller picture of intricate brain functions, like sensory processing, than can be obtained from studying single experiments in isolation. 

\subsubsection*{Challenges \& Concerns}

Extracting new scientific knowledge from new experimental models and increasingly complex datasets also presents many challenges. In particular, it requires mathematical and computational frameworks that can handle the nonlinearity, non-stationarity, and high dimensionality inherent in neural data. Current models often fall short of capturing the full complexity of neural dynamics. Two primary avenues are available for addressing this problem. On the one hand, improved analytical models are needed to describe data in explicitly interpretable ways. On the other hand, numerical methods allow information to be more flexibly extracted from data, though this may be at the expense of rigorous interpretability. A combination of these approaches is likely needed to ensure that complex datasets are analyzed to their full potential.

The creation of digital twins of neural systems provides an interesting opportunity for improving analytical frameworks. A digital twin, in neuroscience, is an artificial system designed to faithfully mimic the dynamics of a specific neural system. In real brains, due to technical limitations, only a small portion of neural activity can be recorded at one time. In contrast, the behavior of a digital twin can be extensively and continuously monitored. Thus, a digital twin that has been well designed and trained to mimic a neural system, likely using machine learning, can generate a considerable amount of data that may, in turn, be ideal for constraining analytical models of the system under study. Certainly, for efforts like these to be successful, collaborations across fields of expertise are needed. Only then can we ensure that the rich datasets being collected are used to their full potential to constrain data-driven models and generate new insights about the brain.

\subsection{\label{sec2:ses2}Neuroscience Toolkits}
Innovative neuroscience toolkits are being developed to ensure cutting-edge tools in neurophysiology research are readily accessible and usable for the broader community. Such toolkits now span the full spectrum of the data lifecycle, with applications to quantifying animal behavior, pre-processing electrophysiology data, interfacing with large-scale data infrastructures, and deploying advanced analysis algorithms, amongst many others. By increasing access to important technical tools, these toolkits help democratize science and facilitate collaborative efforts across the community.

\subsubsection*{Advances \& Examples}
Pre-processing behavioral data to extract relevant information on pose and motion can be a very laborious process, if done manually. Sophisticated machine learning algorithms for motion capture and pose estimation have revolutionized this process, allowing users to automatically extract behavioral information from large data streams. Using DeepLabCut\footnote{For more information on DeepLabCut, see \url{https://github.com/DeepLabCut}.}, for example, users need only manually label a small subset of their dataset to fine-tune the pre-trained pose estimation algorithm to their specific needs and the specific properties of their data. The fine-tuned algorithm can then be deployed on the entire dataset \citep{Mathis2018}. The automation of this process has resulted in a much greater availability of behaviorally labeled neural data. This, in turn, has paved the way for tools like CEBRA\footnote{For more information on Cebra, see \url{https://cebra.ai/}.}, which also uses deep learning, but to identify joint embeddings of behavioral and neural data \citep{Schneider2023}. For both DeepLabCut and CEBRA, operational resources have also been deployed to ensure that these tools are designed, shared and maintained for long-term adoption, community integration, and the democratization of scientific inquiry. Relatedly, MoSeq (Motion Sequencing)\footnote{For more information on MoSeq, see \url{https://dattalab.github.io/moseq2-website}.} is an algorithm that automates the extraction of behavioral data from three-dimensional videos of freely behaving animals captured using depth cameras. By decomposing continuous behavior into distinct ``syllables'' and constructing behavioral state maps, MoSeq provides an unsupervised framework for understanding the sequential and repetitive structure of animal behavior \citep{Wiltschko2020}. This tool can be used, for example, to characterize how different external perturbations—from drug effects to environmental changes—affect behavior over time.
 
There is also a great need for precise and reproducible pre-processing and standardization of electrophysiological data. Accurate signal extraction is hindered by variable levels background noise, differences in acquisition systems, high processing needs for large data volumes, and the complex relationship between intracellular neuronal excitability and extracellular signatures. Even differences in the data formats yielded by acquisition systems can make it difficult to ensure that key pre-processing steps like spike sorting are deployed consistently across experiments. SpikeInterface\footnote{For more information on SpikeInterface, see \url{https://github.com/SpikeInterface}.} aims to address several of these problems by unifying diverse spike sorting algorithms and pre‐processing routines into one Python package, enabling users to compare the outputs of different routines and use consensus metrics to identify reliable units and spikes. It allows for different computational backends, including cloud-based ones, and can be used for data compression to reduce file sizes and for creating web-based, shareable visualizations for quality control and manual curation \citep{Buccino2020,Buccino2023}. Community engagement through active feedback is a core component of SpikeInterface's development, ensuring that the tool continues to evolve to address current and new challenges in electrophysiology research.

Machine learning-based approaches, like those underlying DeepLabCut, CEBRA, and MoSeq, are critical to improving data processing, as they can better adapt to a dataset's specific properties while also being efficiently deployed on large data volumes. Such algorithms have also been created for the detection of neural graphoelements-distinct waveforms or patterns observed in electroencephalography recordings. The detection of sharp wave‐ripples (SWR) in hippocampal recordings has traditionally been done using spectral methods which tend to generalize poorly across brain areas and neurological pathologies, and can bias the types of events that are detected. Machine learning techniques enable more flexible detection of SWRs, but are often tested only on a limited set of datasets. rippl-AI\footnote{For more information on rippl-AI, see \url{https://github.com/PridaLab/rippl-AI}.} brings together in a single toolbox the five best performing algorithms developed in the context of an SWR-detection hackathon. Users can deploy all algorithms at once to obtain aggregated results, and also fine-tune them on their own data, as needed \citep{Navas-Olive2022,Navas-Olive2024}. This project demonstrates how community-based approaches to toolkit development can yield robust and widely applicable tools.

Notably, although the continued creation of new datasets, tools and models is of great value to the neuroscience research community, it can also lead to information overload. The EBRAINS (European Brain ReseArch INfrastructureS) Knowledge Graph\footnote{For more information on EBRAINS, see \url{https://www.ebrains.eu/}.} is a large-scale data integration platform that emerged from the Human Brain Project and tackles this problem by uniting experimental data, computational models, and software tools within a  semantic, linked data framework \citep{Appukuttan2022,Bologna2022}. In so doing, the initiative aims to enhance data discoverability and interoperability. EBRAINS also exemplifies a structural and cultural shift in research toward more collaborative and community-oriented science.

Together, these advances illustrate how longstanding technical challenges can be addressed by engaging members of the research community in developing cutting-edge tools, and investing in their deployment in user-friendly toolkits. This process critically democratizes access to state-of-the-art methodologies and creates a more interconnected, transparent, and sustainable research ecosystem.
 
\subsubsection*{Challenges \& Concerns}
There are many challenges to ensuring that digital resources, like data processing pipelines and analysis tools, remain truly reusable over time. When a digital resource is first shared, its reusability can be maximized by ensuring it is well documented and compatible with other widely used tools. Ensuring that it complies with relevant standardization practices can also greatly enhance its adoption by the research community. Meeting these expectations generally requires significant dedicated effort. In addition, preserving a digital resource's scientific value over time requires continually maintaining and updating it, fostering community engagement, and providing reliable technical support. Open-source software is generally best maintained when processes are streamlined and automated, to the degree possible, and comprehensive testing is deployed. This is necessary, for example, to ensure that a digital resource produces reproducible results across different computational environments. 

Meeting these standards requires time and technical expertise that laboratory members often do not have. Even if they do, transitions in project leadership—often due to the departure of PhD students or postdocs—can disrupt this process, and lead to an otherwise very useful resource becoming obsolete and unusable. The broader community of users, from experimentalists to computational researchers, must also be actively engaged to ensure that digital resources are refined and extended to new and relevant applications. However, facilitating community feedback and contributions can also be very time-consuming. 

Overall, there are many challenges to ensuring the reusability of digital resources over the long term. Broadly, they are likely best addressed through the hiring of dedicated software development and maintenance personnel. This, in turn, requires stable, on-going funding. Adapting current research funding models to allow for such roles will be critical to ensuring that the digital resources produced by research groups continue to flourish.

\begin{table*}[!htbp]
  \caption{Neuroscience toolkits presented or discussed at ODIN 2023.}
    \def\arraystretch{1.5}
    \setlength{\tabcolsep}{3pt}
    \centering
    \begin{tabular}{ | c | c | c | }
    \hline
      \textbf{Resource} & \textbf{Website} & \textbf{Tags} \\
      \hline
      DANDI Archive & \url{https://dandiarchive.org/} & data repository \\ 
      \hline
      EBRAINS & \url{https://search.kg.ebrains.eu} & dataset search, knowledge graph, web app \\
      \hline
      Brain-Score & \url{https://www.brain-score.org/} & dataset search, benchmarks, model evaluation, web app \\
      \hline
      DataJoint & \url{https://datajoint.com/} & data management, database, SQL, Python \\
      \hline
      SpyGlass & \url{https://github.com/LorenFrankLab/spyglass} & data management, database, Python \\
      \hline
      Dendro & \url{https://github.com/flatironinstitute/dendro} & cloud computing, web app \\
      \hline
      Neurosift & \url{https://neurosift.app} & visualization, dataset exploration, web app \\
      \hline
      NWB GUIDE & \url{https://github.com/NeurodataWithoutBorders/nwb-guide} & data format conversion, desktop app \\
      \hline
      NeuroConv & \url{https://github.com/catalystneuro/neuroconv} & data format conversion, Python \\
      \hline
      Neo & \url{https://github.com/NeuralEnsemble/python-neo} & data format reading, Python \\
      \hline
      SpikeInterface & \url{https://github.com/SpikeInterface/spikeinterface} & spike sorting, electrophysiology, Python  \\ 
      \hline
      rippl-AI & \url{https://github.com/PridaLab/rippl-AI} & SWR detection, electrophysiology, Python \\
      \hline
      OptiNiSt & \url{https://github.com/oist/optinist} & ROI segmentation, optical physiology, desktop app  \\ 
      \hline
      Caiman & \url{https://github.com/flatironinstitute/CaImAn} & ROI segmentation, optical physiology, Python  \\ 
      \hline
      EXTRACT & \url{https://github.com/schnitzer-lab/EXTRACT-public} & ROI segmentation, optical physiology, MATLAB  \\ 
      \hline
      suite2p & \url{https://github.com/MouseLand/suite2p} & ROI segmentation, optical physiology, Python  \\ 
      \hline
      DeepLabCut & \url{https://github.com/DeepLabCut/DeepLabCut} & pose estimation, behavior, Python \& desktop app \\ 
      \hline
      Lightning Pose & \url{https://github.com/danbider/lightning-pose} & pose estimation, behavior, Python \\ 
      \hline
      SLEAP & \url{https://github.com/talmolab/sleap} & pose estimation, behavior, Python \& desktop app \\ 
      \hline
      VAME & \url{https://github.com/LINCellularNeuroscience/VAME} & pose estimation, behavior, Python \\ 
      \hline
      MoSeq & \url{https://github.com/dattalab/moseq2-app} & video sequencing, behavior, Python \\ 
      \hline
      CEBRA & \url{https://github.com/AdaptiveMotorControlLab/CEBRA} & data analysis, latent space, behavior, Python \\ 
      \hline
      Pynapple & \url{https://github.com/pynapple-org/pynapple} & data analysis, Python \\ 
    \hline
  \end{tabular}
  \label{tab:toolkits}
\end{table*}

\subsection{\label{sec2:ses3}Modeling and Benchmarking}
On the modeling side of neuroscience, integrative approaches are also increasingly being embraced, as exemplified by toolkits like SpikeInterface and rippl-AI described above. Such approaches aim to bring together single, isolated models under a shared infrastructure, allowing them to be compared and even used in conjunction with one another. Successful integration requires the development of robust benchmarks against which existing and new models can be compared. This, in turn, requires high-quality, large-scale data covering a wide breadth of tasks and datasets to be available for model constraining and benchmarking. Overall, the task of moving toward more integrated computational neuroscience will require the community to bridge gaps between disparate datasets, metrics and models.

\subsubsection*{Advances \& Examples}

Brain-Score\footnote{For more information on Brain-Score, see \url{https://www.brain-score.org/}.} is a platform for integrative benchmarking of models of the brain. By connecting large but previously disconnected datasets, Brain-Score creates a common ground for evaluating models on a wide range of neural and behavioral tasks, and identifying classes of models that best recapitulate the brain's functions \citep{Schrimpf2018, Schrimpf2020}. A framework like this not only provides robust constraints for model development and a mechanism for rapid model screening, but also offers an avenue for predicting experimental outcomes and optimizing data collection strategies \citep{Tuckute2024,Schrimpf2024,Bashivan2019}. Ensuring that biophysical and anatomical constraints are appropriately integrated into computational model design and training can also greatly improve their ability to capture real neural data. For example, traditional neural networks trained on detailed physiological data can learn to recapitulate flexible neural encoding of movement and to generalize across conditions, as the complexity of the data they are trained on requires them to identify robust solutions \citep{Saxena2022,Almani2022,Almani2024}. This illustrates the importance of ensuring that high-quality data covering a wide breadth of task conditions is available for training and evaluating computational models.

New metrics and baselines are also needed to ensure computational models are robustly evaluated against real data. Research has shown that despite extensive remapping of place cells, animals can rapidly reapply learned responses across different environments. Manifold analyses deployed on this data show that task-specific representations can remain stable even through remapping, pointing to geometrical consistency across tasks \citep{Wirtshafter2024}. Such findings can provide concrete, and importantly non-trivial, targets and constraints for computational models. Separately, drawing on principles from shape theory, ``shape metrics'' enable high-dimensional neural representations to be compared across individuals or across models \citep{Williams2021,Duong2023,Pospisil2024,Harvey2023}. The variability measured can in turn be related to behavioral variability to determine how closely intertwined the two are in the specific brain region under study. Such metrics can provide also concrete targets for computational models, ensuring that they capture the patterns observed both within and across the brains of real subjects. Notably, these examples also highlight the value of using metrics derived from unsupervised analyses, in addition to supervised ones, to evaluate models.

Together, these efforts demonstrate the importance of developing integrated benchmarks evaluated on high-quality and varied datasets with robust metrics, and ensuring that benchmarking tools can be efficiently deployed on new models. Advances in this direction are critical for ensuring that computational research in neuroscience is well-integrated and grounded in real data.

\subsubsection*{Challenges \& Concerns}
There are many challenges to integrating the field of computational neuroscience. First, many fields of research, such as the study of motor function, currently lack robust benchmarks against which models can be evaluated. Ideally, individual benchmarks should not be used as sole criteria for model selection. A wide range of benchmarks used in parallel can provide a more global and representative perspective on a model's performance. In fields for which many different metrics and analysis techniques exist, like dimensionality reduction and manifold analysis, the wealth of options can also present a challenge. Researchers less familiar with the theoretical grounding of individual techniques may struggle to identify the approach most relevant to their dataset and research question. Well-documented toolkits that bring together a variety of related metrics and analysis techniques, and include clear usage guidance are needed to ensure that these tools are used correctly and in the appropriate setting.

Furthermore, benchmarks are only as robust as the data on which they rely. Thus, as mentioned above, high-quality and varied data is necessary for successful and reliable benchmarking. Thorough quality assurance procedures and rich metadata are indispensable to ensure that datasets used for benchmarking and for model training accurately reflect the neural activity and behaviors under study. Lastly, even once robust benchmarks are created, bringing together complex models built using different architectures and even software under a shared benchmarking platform, as Brain-Score has done for vision and language models, requires considerable time and effort. Community investment and sustained collaborations, as well as a willingness by the community to adopt standardized methodologies, are critical to the success of such projects.

\begin{table*}[hbt]
\def\arraystretch{1.5}
\setlength{\tabcolsep}{3pt}
\centering
\caption{BRAIN Initiative data archives.} 
\vspace{-1.2em}
\captionof*{table}{These archives are generally public access, although some house restricted datasets. Most of these archives also allow embargoes, i.e., restricted access for a fixed period of time after initial publication.}
\begin{tabular}{|p{0.25\textwidth}|l|p{0.24\textwidth}|p{0.25\textwidth}|}
\hline
\textbf{Archive} & \textbf{Link} & \textbf{Datatypes} & \textbf{Access Restrictions} \\
\hline
\textbf{BIL} \newline (\textbf{B}rain \textbf{I}maging \textbf{L}ibrary) & \url{https://www.brainimagelibrary.org/} & Confocal microscopy brain \newline imaging & Some restricted datasets \\
\hline
\textbf{bossDB} \newline (\textbf{B}lock and \textbf{O}bject \newline \textbf{S}torage \textbf{S}ervice \textbf{D}ata\textbf{b}ase) & \url{https://bossdb.org/} & Electron microscopy and \newline x-ray microtomography & Public \\
\hline
\textbf{DABI} \newline (\textbf{D}ata \textbf{A}rchive for the \textbf{B}RAIN \textbf{I}nitiative) & \url{https://dabi.loni.usc.edu/} & Invasive human \newline neurophysiology & Some restricted datasets \newline and requires registration \\
\hline
\textbf{DANDI} \newline (\textbf{D}istributed \textbf{A}rchives \newline for \textbf{N}europhysiology \textbf{D}ata \textbf{I}ntegration) & \url{https://www.dandiarchive.org/} & Cellular, systems, and \newline behavioral neurophysiology & Public \\
\hline
\textbf{NEMAR} \newline (\textbf{N}euro\textbf{e}lectro\textbf{m}agnetic Data \textbf{A}rchive \newline and Tools \textbf{R}esource) & \url{https://nemar.org/} & Electroencephalogram (EEG) and \newline magnetoencephalography (MEG) & Public \\
\hline
\textbf{NeMO} \newline (\textbf{Ne}uroscience \textbf{M}ulti-\textbf{O}mic \newline Data Archive) & \url{https://nemoarchive.org/} & Multi-omics & Some restricted datasets \\
\hline
\textbf{OpenNeuro} & \url{https://openneuro.org/} & Magnetic resonance \newline imaging (MRI) and other \newline types of neuroimaging & Public \\
\hline
\end{tabular}
\label{tab:brain_archives}
\end{table*}

\begin{table*}[hbt]
\def\arraystretch{1.5}
\setlength{\tabcolsep}{3pt}
\centering
\caption{Generic archives that contain some neurophysiology data.}
\vspace{-1.2em}
\captionof*{table}{All of these are public access.}
\label{tab:additional_archives}
\begin{tabular}{|p{0.25\textwidth}|l|p{0.26\textwidth}|}
\hline
\textbf{Archive} & \textbf{Link} & \textbf{Datatype} \\
\hline
\textbf{Brain/MINDS Data Portal} \newline (Japan’s Brain Mapping Project) & \url{https://dataportal.brainminds.jp/} & Includes marmoset structural \newline and functional physiological data \\
\hline
\textbf{Brain-Score} & \url{https://www.brain-score.org/} & Neurophysiology data \\
\hline
\textbf{CRCNS} \newline (\textbf{C}ollaborative \textbf{R}esearch \newline in \textbf{C}omputational \textbf{N}euro\textbf{s}cience) & \url{https://crcns.org/} & Neurophysiology data \\
\hline
\textbf{Dryad} & \url{https://datadryad.org/} & General research data \\
\hline
\textbf{EBRAINS} & \url{https://ebrains.eu/} & Various types of neuroscience data \\
\hline
\textbf{Figshare} & \url{https://figshare.com/} & General research data \\
\hline
\textbf{G-NODE} \newline (\textbf{G}erman Neuroinformatics \textbf{Node}) & \url{https://gin.g-node.org/} & Neurophysiology data \\
\hline
\textbf{Zenodo} & \url{https://zenodo.org/} & General research data \\
\hline
\end{tabular}
\label{tab:archives}
\end{table*}

\section{\label{sec3}Key Technical Hurdles and Solutions}

Having summarized many of the recent advancements in neurophysiology and justified their value to the field, we next examine in more detail specific challenges preventing widespread adoption of the technologies and practices. We also describe some possible solutions gleaned from other scientific areas that have experienced similar difficulties.

\subsection{Data Standards}
The United States' government's repository of Federal Enterprise Data Resources defines a ``data standard'' as a ``technical specification that describes how data should be stored or exchanged for the consistent collection and interoperability of that data across different systems, sources, and users.'' Such a specification comprises components like datatype, identifiers, vocabulary, schema, format, and API\footnote{For more information on data standards, see \url{https://resources.data.gov/standards/concepts/#data-standard}.} In this section, we discuss challenges to consistency and completeness when using data standards in neurophysiology.

\paragraph{\textbf{Data Standard Strictness.}}
How strictly or broadly the components of a data standard are defined greatly influences its applicability, usability and flexibility. For example, a very narrowly defined data standard might be easy to use, but applicable only to a handful of datasets. In contrast, the NWB standard aims to be applicable to the full range of neurophysiology datasets. Questions persist, however, about whether certain components, such as metadata vocabulary, and supported storage formats and APIs should be more narrowly specified to improve consistency and reusability. Existing perspectives generally favor providing recommended best practices and sensible defaults rather than imposing stringent rules, ensuring the NWB standard can adapt to the diverse needs found in neurophysiology research.

\paragraph{\textbf{Ontologies.}}
Establishing links between descriptive terms used in neurophysiology and standardized ontologies\footnote{An ontology provides a structured framework for organizing and connecting descriptive terms, facilitating better understanding and communication within a field.} can greatly improve clarity and communication. Associating a brain area studied in an experiment with its corresponding region in a widely used atlas is one example. Although resources like the National Center for Biotechnology Information (NCBI) Taxonomy, the Mouse Genome Informatics (MGI) database, and the Neuroscience Information Framework (NIF) Standard Ontology are available, they are often underutilized due to the additional effort required to consult them when entering identifiers into metadata. One proposed strategy is to develop interfaces that provide default options or infer metadata directly from the data, thereby simplifying and standardizing the metadata entry process. As above, balancing strict enforcement of accepted ontologies with the flexibility that keeps data usable remains an ongoing challenge.

\paragraph{\textbf{Standardizing Initial Data Recording.}}
To promote the adoption of the data standards, data acquisition systems could be enabled to write raw data directly into standards like NWB. While this would greatly alleviate pressure on individual researchers and laboratories to handle data conversion, a direct conversion approach does present certain challenges. A first challenge concerns the comprehensiveness of metadata. Metadata includes information about data collection, experimental design, and subject details, all of which provide essential context for experimental data. However, many of these details are often collated after data acquisition begins from disparate sources like laboratory notebooks. As a result, the NWB files created during acquisition may initially lack compliance with the standard’s own metadata requirements. Extending the ecosystem of NWB-related tools to help researchers enter all available metadata at time of data collection, and add new information as it becomes available could alleviate this problem. Metadata entry tools could also be used to encourage researchers to record a broader range of information, including environmental factors (e.g., temperature, humidity, luminance) which is often omitted when recording experiment variables.

A second challenge is the problem of data stream alignment. It is very common in neurophysiology experiments for multiple time-based data streams, like neural and behavioral activity, to be collected in parallel, aligned to separate clocks. In fact, when using devices like the Neuropixels probe, data from various channels are not sampled exactly simultaneously, but are instead slightly offset. The NWB standard requires data streams to be aligned to a common clock within a single NWB file. This alignment process is usually done offline after data collection, often using custom scripts, and can require user intervention to handle irregularities caused, for example, by hardware failures. The NWB standard could be modified to allow unaligned raw data streams to be recorded in their native clocks along with synchronizing pulse data. However, the wide variety in experimental setups and lack of universally accepted methods for sending synchronizing signals creates a considerable challenge for expanding data standards to accommodate idiosyncratically recorded raw data streams. Even if raw data streams could effectively be directly recorded in NWB files, it would be critical for data collectors to ensure aligned data streams are added to the files or to new files created from scratch before they are shared. Indeed, although some common alignment methods exist, such as ``tshift'' for synchronizing Neuropixels channels (as is done in SpikeGLX \citep{SpikeGlxDocs2025}), many alignment protocols are laboratory specific. Sharing NWB files with only unaligned raw data streams would thus seriously impede the ability of end users to properly use the data, thus defeating one of the core goals of data standardization. Thus, although enabling data acquisition systems to write raw data directly into standards like NWB could be tremendously useful to the field, key obstacles remain to be addressed to create a sufficiently feasible and flexible solution.

\paragraph{\textbf{Data Curation.}}
As mentioned above, experimental information like session or subject exclusions is frequently stored in laboratory notebooks or separate databases, yet integrating it into the data sharing process is vital for a complete understanding of dataset usage and proper interpretation of analyses. Since such annotations are often highly context-specific and free-form, determining how to embed them within data standards while maintaining sufficient flexibility for various experimental designs remains the object of ongoing discussions. These considerations again reflect the broader issue of standardizing experimental metadata in neurophysiology, where efforts must balance strict requirements with the fluidity needed to accommodate a spectrum of research approaches.

\paragraph{\textbf{Provenance Storage.}}
Determining where to store provenance information—inputs, settings, and outputs from computational analyses—is another key consideration. Some standards and data management systems record provenance in separate files from the data. ALPACA, for instance, stores these data in Resource Description Framework (RDF) files \citep{Kohler2024}, while DataJoint uses a database-backed processing pipeline \citep{Yatsenko2018}. In contrast, data standards such as NWB aim to store all information related to an experiment within a single file. Provenance information could be integrated into the NWB standard by, for example, recording information similar to what is stored in RDF files directly into single NWB files. However, this would require adding another layer of complexity to creating NWB files and maintaining the NWB standard. Thus, leaving provenance management to external systems might be more practical.

\subsection{Common Infrastructure and Computational Reproducibility}
Neurophysiology data processing and analyses often depend on software packages with specific environmental and operating system requirements. Versions of these packages, their dependencies, and their installation environments are seldom recorded alongside analysis results. This omission complicates both the reproduction of results and the assessment of software bugs’ potential impact on data analysis outcomes.

\paragraph{\textbf{Containerization as a Solution.}}
Containerization through tools like Docker presents a means of addressing reproducibility hurdles. A Docker container packages an application together with its dependencies, ensuring consistent behavior across different computing environments. Documenting this build process in a Dockerfile captures every command involved in creating the application’s environment and installing its required packages, facilitating the containerization process \citep{Nust2020}. By sharing not only the data but also the entire computational setup used to process that data, researchers can substantially enhance the reproducibility of scientific findings. Further gains in computational reproducibility may also emerge from integrating Docker containers within data standards. The NWB standard includes an optional ``source-script'' field for storing a Uniform Resource Identifier (URI) linking to a container image and the scripts used for analysis, allowing other researchers to replicate a computational environment. This option could be expanded to allow individual data streams within a single file to be linked to distinct scripts or container often operating on separate clocks, since NWB often holds multiple processed data streams produced by different recordings and analyses.

\paragraph{\textbf{Harnessing AI.}}
Integrating artificial intelligence (AI) into neurophysiology research, particularly in the context of leveraging open data, comes with promise but also challenges. While exploring strategies for maximizing the benefits of these technologies, we must acknowledge the complexity, scale, and peculiarities of neurophysiology data. Several underlying considerations will shape the role of AI in neurophysiology research. One is determining whether the scientific community is ready to depend on LLMs for data analysis given the opacity of their inner workings. A second is determining how doctoral researchers should balance time in developing computational expertise and leaning more on AI tools and specialists. A third is clarifying how much foundational knowledge is necessary to avoid misuse of AI techniques and misinterpretation of results by neuroscience researchers with non-computational backgrounds.

As the community moves to adopt more AI methodologies, three primary barriers nonetheless hinder their broad application to neurophysiology data:

\begin{itemize}[noitemsep,nolistsep]
    \item \textbf{Experiment Diversity and Dataset Quality.} The heterogeneity of experimental designs makes it difficult to generate consistent data for meta‐analyses, highlighting the need for `AI-ready datasets'. The challenge is that these datasets must be exceptionally large in sample size, well‐organized, and mutually consistent in order to support the training and validation of effective AI models \citep{Deng2012, Chen2021}.
    \item \textbf{Common Vocabulary for Neural Patterns.} The lack of consensus on a common vocabulary to describe neural patterns (such as ripples, bursts, avalanches, etc.) impedes the generalizability of AI methods.  A shared terminology is vital for effective communication throughout the community. Shared terminologies have been more widely adopted, for example, for describing the classes of neural computations based on underlying low-dimensional manifolds \citep{Vyas2020}, but are still lacking or incomplete when it comes to describing dendritic, neuronal, or microcircuit activity. As a result, although automated methods, such as sequential spectral density approaches or deep neural networks, have been proposed for spindle detection \citep{Dehghani2011,Kaulen2022}, sharp-wave ripple analysis \citep{Liu2022,Ramirez2015,Navas-Olive2022}, and cell-type classification based on spiking waveforms \citep{Bartho2004,Peyrache2012,Telelenczuk2017}, their applicability and reach remain limited due to a lack of universal agreement on conventions.
    \item \textbf{Annotations.} It is currently uncommon for neurophysiology datasets to receive annotation beyond the original experimenter(s). By inviting external users to contribute annotations to datasets across all archives, unexpected meta-patterns may be discovered and overall ``AI-readiness'' increased \citep{Raddick2009,Banerji2010}. Common standards such as BAABL \citep{Ly2024} and Hierarchical Event Descriptors (HED) \citep{Bigdely2016} could also be used to standardize and improve data annotations for even greater reusability.
\end{itemize}

\paragraph{\textbf{Data Quality and Benchmarks.}}
Relatedly, it is important for the community to be able to assess the usability of a dataset prior to training an AI model. Reliable metrics are needed to evaluate and transparently report the quality of both raw and processed data sources. These can provide helpful feedback for improving future experiments. The MRI Quality Control tool provides a compelling example \citep{Esteban2017}. Using such metrics as benchmarks, public leaderboards for model training can be hosted to incentivize competitive groups to achieve top ranks \citep{Schrimpf2018, Schrimpf2020}, thus accelerating progress in research.

Such strategies point to a strong need for community‐driven tools and collaborative approaches, which will become increasingly important as AI plays a larger role in neurophysiology research. Further developments in each of these areas are expected to improve the quality of open data over time.

\paragraph{\textbf{Community Engagement and Collaboration.}}
Finally, several obstacles exist when it comes to engaging the neurophysiology community in using new tools, as well as in collaborating on projects. First, despite the potential of new tools like containerization and AI methodologies, widespread adoption requires broader user familiarity and more accessible infrastructure for the neuroscience community. Researchers must also be supported in managing the costs associated with long-term archiving and maintaining of outputs like container images and trained neural networks. Addressing these obstacles is vital to ensuring the viability of improved reproducibility and analysis methods for the broader neurophysiology community. Second, many laboratories have and continue to develop their own frameworks and terminologies tailored to specific experimental protocols. These lab-specific rules and approaches can lead to confusion and limit collaboration. While this diversity offers certain advantages, establishing a clearer common ground and shared tools would facilitate communication, reduce misunderstandings, and support more consistent use of effective methodologies.

\section{\label{sec4} Looking ahead}

In this section, we discuss recommendations for strengthening the open science community moving forward. Specifically, we will discuss the importance of building community and of appropriately harnessing large language models (LLMs) as a new and potentially revolutionary tool. We also list current community needs, as well as recommendations for both the practicing neuroscientist and the neuroscience science community as a whole.

\subsection{\label{sec4:item0}Building Communities}

\paragraph{\textbf{Building a Community.}} 
Open science thrives in a well-supported ecosystem where community-based governance and communication can flourish \citep{Saunders2022}. The nascent but growing ODIN community will require robust mechanisms for dialogue and self-regulation, ideally emerging organically from within the community itself. A prime example of this model is Wikipedia, which thrives under self-imposed rules and a transparent decision-making process. Unlike transient tools like team communication platforms, a wiki provides a durable, public, and cumulative resource for community discourse \citep{Kelder2012}. Engaging in quality discussions and integrating these alongside the data itself will ensure accessibility and transparency for the wider public.

\paragraph{\textbf{Regular Meetings.}} 
The enthusiasm shared during this symposium suggests a strong desire for it to continue on a regular basis. These meetings are envisioned as key catalysts for fostering a robust ODIN community, and drawing together diverse voices from across the neurophysiology and systems neuroscience spectrum. By maintaining open communication channels and featuring varied perspectives, we hope to enrich our collective knowledge. In addition, we hope that continuing to share these talks on widely-used and open video sharing platforms will ensure broad, public accessibility and engagement.

\subsection{\label{sec4:item1}Harnessing Large Language Models (LLMs)}
The advent of advanced LLMs such as \textit{OntoGPT} \citep{Caufield2023} and \textit{BrainGPT} \citep{Luo2024} heralds a transformative shift in how scientific information can be processed, understood, and utilized. These models have demonstrated a remarkable ability to distill and predict complex patterns from vast datasets, suggesting a potential role in enhancing user interaction with neuroscientific databases. AmadeusGPT showcases an innovative application in this direction, using LLMs to convert natural language descriptions of animal behaviors into executable analysis code, thereby facilitating interactive behavioral research, and increasing its accessibility \citep{Ye2023}. Tools like these exemplify the potential of LLMs to help bridge gaps between complex biological knowledge and expertise in computational analysis, enhancing scientists' ability to access and analyze neuroscience data.

LLMs also have the potential to help researchers engage more effectively with existing scientific knowledge, for example through enhanced literature searches and dynamic knowledge base augmentation via scientific journal content distillation. For example, BrainGPT has been specifically trained to anticipate the outcomes of neuroscience experiments by ingesting extensive portions of the neuroscientific literature. Its proficiency, as demonstrated by the BrainBench benchmark, was shown to surpass that of human experts in distinguishing between true experiment results and modified abstracts \citep{Luo2024}. This suggests a potential use for LLMs in inductive reasoning \citep{Wang2023} where an LLM trained on open data is harnessed for hypothesis generation and experiment planning. Capabilities like these also point to a future where LLMs could be used to reliably navigate and summarize existing scientific knowledge. It is important to note, however, that LLMs are subject to hallucinations. For this reason, BrainGPT is not currently enabled to perform this type of task \citep{Huang2023}, and this potential use remains a matter of conjecture for the moment. However, such models could be used to refine literature search mechanisms, enabling researchers to rapidly locate relevant studies and datasets. By processing queries using LLMs trained on the latest research and reviews, search engines could offer more contextually aware search results, reducing the time spent on literature reviews and increasing the relevance of the information retrieved.

OntoGPT's approach to enhancing knowledge bases through natural language processing highlights another very promising application for LLMs \citep{Caufield2023}. By helping construct and refine knowledge bases, LLMs can facilitate more accurate and dynamic querying of complex data structures. When it comes to managing extensive open neuroscientific data repositories, integrating LLMs could dramatically improve the precision and scope of data retrieval processes, enabling researchers to generate interconnected insights from disparate datasets. They could also be used to ensure that data curation and query management follow standard nomenclature and are consistent with open neuroscience practices. For example, LLMs can assist in linking new data entries with existing ontologies and suggest updates to improve the comprehensiveness and utility of a database.

Thus, as LLMs evolve, we should be able to leverage them not only to manage and query existing data, but also to anticipate and prepare for future research developments. Continuous updates to LLM training sets and algorithms and fine-tuning, using methods like LoRA \citep{Hu2021} and Retrieval Augmented Generation (RAG) \citep{Lewis2021} techniques, will be essential to maintaining their effectiveness and relevance to the neuroscientific context. Overall, incorporating LLMs into open data in neuroscience promises to enhance utility as a dynamic, forward-looking tool that not only serves current user needs, but also adapts to and anticipates future scientific challenges.

\subsection{Community Needs}
For open science to advance in neuroscience, it is critical to understand and address the needs of the community. Table \ref{tab:community_needs} summarizes key community needs identified during the symposium. Most broadly, these include a need for improved resources and tools, including educational ones, that make open science practices easier to adopt for users of all backgrounds. However, another key need is for changes to funding and incentives that ensure users can invest the time and effort required to adopt open science practices. In particular, incentives that recognize and reward data sharing, collaborative model development, and interdisciplinary efforts are needed to build a truly open and innovative neuroscience research community.

\begin{table*}[ht]
\centering
\caption{Community needs and actions for advancing open science.}
\def\arraystretch{1.5}
\setlength{\tabcolsep}{3pt}
\begin{tabular}{|p{0.16\linewidth}|p{0.55\linewidth}|p{0.2\linewidth}|}
\hline
\textbf{Category} & \textbf{Actions} & \textbf{Key Concepts} \\ \hline
Guidance & 
\vspace{-\baselineskip}
\begin{itemize}[nosep,topsep=0.5em,left=0.5em]
    \item Provide community guidance on sharing methodologies, datatypes (raw, processed).
    \item Standardize required and recommended metadata types.
    \item Select and unify ontologies for metadata standardization.
    \item Define essential provenance information for shared data.
    \vspace*{-\baselineskip}
\end{itemize} & 
Provenance, shared methodologies, standardized metadata, unified ontologies \\ \hline
Tool Development & 
\vspace{-\baselineskip}
\begin{itemize}[nosep,topsep=0.5em,left=0.5em]
    \item Enhance tools for data compression, conversion, sharing, and analysis.
    \item Develop cloud-based data access and analysis solutions.
    \item Establish benchmarking platforms for model and theory evaluation.
    \item Develop platforms for tool comparison.
    \item Support large-scale data pooling and annotation.
    \item Simplify metadata entry through user-friendly interfaces.
    \item Improve automated metadata capture tools.
    \item Enable on-the-fly data annotation of anomalies during experiments.
    \item Improve ability to detect and filter anomalous data.
    \vspace*{-\baselineskip}
\end{itemize} & 
Cloud solutions, data compression, data pooling, metadata entry, tool benchmarking \\ \hline
Research &
\vspace{-\baselineskip}
\begin{itemize}[nosep,topsep=0.5em,left=0.5em]
    \item Improve models for understanding complex data.
    \item Create benchmarks and metrics for model evaluation.
    \item Develop data quality assurance metrics.
    \item Innovate on automated data labeling for enhanced data reuse.
    \vspace*{-\baselineskip}
\end{itemize} & 
Advanced model zoo, automated data labeling, data quality metrics, model benchmarks \\ \hline
Databases &
\vspace{-\baselineskip}
\begin{itemize}[nosep,topsep=0.5em,left=0.5em]
    \item Maintain centralized databases for datasets, methodologies, and tools.
    \item Facilitate community feedback mechanisms for shared resources.
    \vspace*{-\baselineskip}
\end{itemize} & 
Centralized databases \\ \hline
Knowledge \& Education &
\vspace{-\baselineskip}
\begin{itemize}[nosep,topsep=0.5em,left=0.5em]
    \item Create knowledge graphs for describing entities and their relationships, and for linking disparate databases.
    \item Continue to develop online resources and training for data processing and analysis tools.
    \vspace*{-\baselineskip}
\end{itemize} &
Knowledge graphs, online resources, training workshops \\ \hline
Funding \& Incentives &
\vspace{-\baselineskip}
\begin{itemize}[nosep,topsep=0.5em,left=0.5em]
    \item Support community engagement and multi-laboratory collaborations.
    \item Fund technical personnel for open-source software maintenance.
    \item Fund the creation of core facilities in research institutions that provide centralized technical expertise to individual laboratories. 
    \item Encourage and facilitate adoption of new technologies and open science practices.
    \item Invest in scaling data storage solutions.
    \vspace*{-\baselineskip}
\end{itemize} &
Community engagement, core facilities, multi-laboratory collaboration, open-source support \\ \hline
\end{tabular}
\label{tab:community_needs}
\end{table*}

\subsection{Recommendations for the Practicing Neuroscientist}

As we continue to innovate and advocate for community-wide advancements in open science, practicing neuroscientists have several opportunities to engage with the existing open science practices. This section makes recommendations that span the entire lifespan of a project and can greatly improve the reproducibility and efficiency of one's research. Depending on their projects and access to resources, individual research groups may find certain recommendations more relevant, helpful or feasible to implement than others. We recommend identifying these priorities and approaching the adoption of open science practices incrementally.

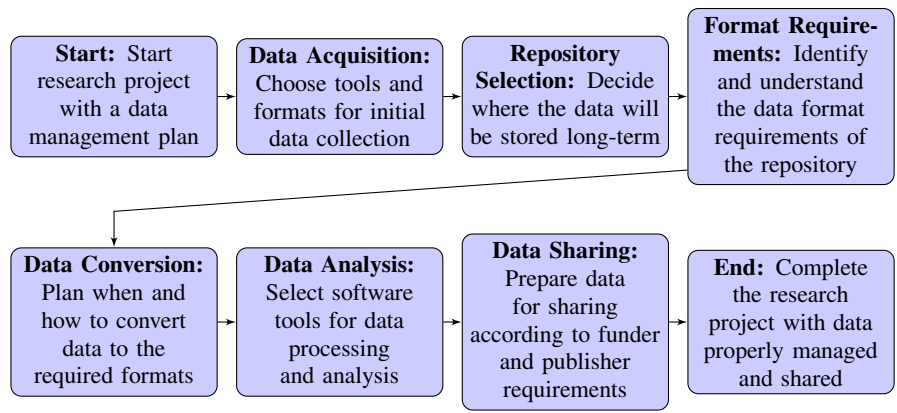
\begin{figure*}[hbt]
\centering
\begin{tikzpicture}[node distance = 4.7cm, auto]
    \tikzstyle{block} = [rectangle, draw, fill=myblue!30!white, 
    text width=11.5em, text centered, rounded corners, minimum height=4em]
    \tikzstyle{line} = [draw, -latex']
    \node [block] (start) {\textbf{Start:} Start research project with a data management plan. \\~\\
    \textit{\textbf{Example:} Plan to collect neurophysiology data, convert to a common standard, store temporarily on an external drive, and long-term in an online repository.}};
    \node [block, right of=start] (acquisition) {\textbf{Data Acquisition:} Choose tools and formats for initial data collection. \\~\\
    \textit{\textbf{Example:} Plan to collect dataset in the proprietary formats exported by the data collection software.}};
    \node [block, right of=acquisition] (repository) {\textbf{Repository Selection:} Decide where the data will be stored long-term. \\~\\
    \textit{\textbf{Example:} Select the DANDI Archive for long-term storage.}};
    \node [block, right of=repository] (format) {\textbf{Format Requirements:} Identify and understand the data standard requirements of the repository. \\~\\
    \textit{\textbf{Example:} Identify NWB as the appropriate format for the datatype and the selected repository.}};
    \node [block, below of=start] (conversion) {\textbf{Data Conversion:} Plan when and how to convert data to the required formats. \\~\\
    \textit{\textbf{Example:} Decide to convert the data after collection, but before analysis. Develop a pipeline to be deployed as data is collected.}};
    \node [block, right of=conversion] (analysis) {\textbf{Data Analysis:} Select software tools for data processing and analysis. \\~\\
    \textit{\textbf{Example:} Plan to use CaImAn for ROI segmentation, DeepLabCut for pose estimation and Pynapple for analysis.}};
    \node [block, right of=analysis] (sharing) {\textbf{Data Sharing:} Prepare data for sharing according to funder and publisher requirements. \\~\\
    \textit{\textbf{Example:} Upload the dataset to the DANDI Archive, and publish it publicly.}};
    \node [block, right of=sharing] (end) {\textbf{End:} Complete the research project with data properly managed and shared. \\~\\
    \textit{\textbf{Example:} Share links to dataset at conferences and talks to encourage others to use it.}};
    \path [line] (start) -- (acquisition);
    \path [line] (acquisition) -- (repository);
    \path [line] (repository) -- (format);
    \path [line] ($(format.south west) + (0,0.2)$) -- ($(conversion.north east) - (0, 0.2)$);
    \path [line] (conversion) -- (analysis);
    \path [line] (analysis) -- (sharing);
    \path [line] (sharing) -- (end);
\end{tikzpicture}
\caption{Data management plan flowchart, with example.}
\label{fig:planFlow}
\end{figure*}

\paragraph{\textbf{Data Management and Sharing Plan.}} It is important to prepare a data management and sharing plan early in the research process. Funders like the NIH and many scientific journals now require open sharing of data collected under their grants and for publication, respectively. Deciding early on which repository to use, understanding its data standard requirements, and planning the workflow from data acquisition to publication can greatly facilitate the process. Adopting standards such as NWB early in data acquisition can also streamline the process and save time by ensuring consistency is maintained through data processing, analysis, and publication. See Fig. \ref{fig:planFlow} for steps of a recommended workflow. Additional recommendations are listed below for optimizing the process of selecting and converting to a data standard:
    \begin{enumerate}
    \item Identify early on the \textbf{repository} where your data will be deposited. Understanding the \textbf{data standard requirements} of the chosen repository and whether it needs to conform to specific standards such as NWB or BIDS will critically shape future steps.
    \item Consider how you will \textbf{manage, process, analyze, and visualize} your data. The software tools you will use may have limitations on the data formats they accept, and the formats they can output.
    \item Plan to \textbf{convert data} to a common standard \textbf{as early as possible} in the data lifecycle: from acquisition through processing, analysis, and up to publication and sharing. Early adoption streamlines workflows, avoids the need to refactor custom code down the road, and enhances the reusability of data, saving time and resources.
    \item \textbf{Automate}, to the extent possible, the process of \textbf{converting your data} into the required standard using tools like NWB GUIDE. If conversion is done at the acquisition step, make sure that the proper metadata is included. If certain post-processing steps are required routines in your laboratory, make sure to track the details of how they were run and include the relevant information during data conversion to standardized formats.
    \end{enumerate}

\begin{figure*}[hbt]
\centering
\begin{tikzpicture}[node distance = 3.5cm, auto]
    \tikzstyle{block} = [rectangle, draw, fill=myblue!30!white, 
    text width=8em, text centered, rounded corners, minimum height=4em]
    \tikzstyle{line} = [draw, -latex']
    \node [block] (start) {\textbf{Start:} Begin your research project with a plan for comprehensive metadata documentation.};
    \node [block, right of=start] (collection) {\textbf{Data Collection:} Record data collection details (including devices and software).};
    \node [block, right of=collection] (processing) {\textbf{Data Processing:} Document any data transformations and analyses, specifying software and versions.};
    \node [block, right of=processing] (alignment) {\textbf{Alignment and Integration:} Detail methods for aligning and integrating multiple data streams and types.};
    \node [block, below of=start, yshift=-0.5cm] (annotation) {\textbf{Anomaly Annotation:} Note any experimental anomalies and their potential impact on the data.};
    \node [block, right of=annotation] (finalization) {\textbf{Metadata Finalization:} Use tools like NWB GUIDE to finalize the metadata documentation, ensuring it is detailed and standardized.};
    \node [block, right of=finalization] (sharing) {\textbf{Data Sharing:} Share data along with metadata in the chosen repository, making it accessible and usable for the broader research community.};
    \node [block, right of=sharing] (end) {\textbf{End:} Complete the project with well-documented, reusable data that can support future research.};
    \path [line] (start) -- (collection);
    \path [line] (collection) -- (processing);
    \path [line] (processing) -- (alignment);
    \path [line] ($(alignment.south west) + (0,0.15)$) -- ($(start.south)!0.5!(annotation.north)$) -- (annotation.north);
    \path [line] (annotation) -- (finalization);
    \path [line] (finalization) -- (sharing);
    \path [line] (sharing) -- (end);
\end{tikzpicture}
\caption{Documentation/Metadata flowchart.}
\label{fig:docFlow}
\end{figure*}
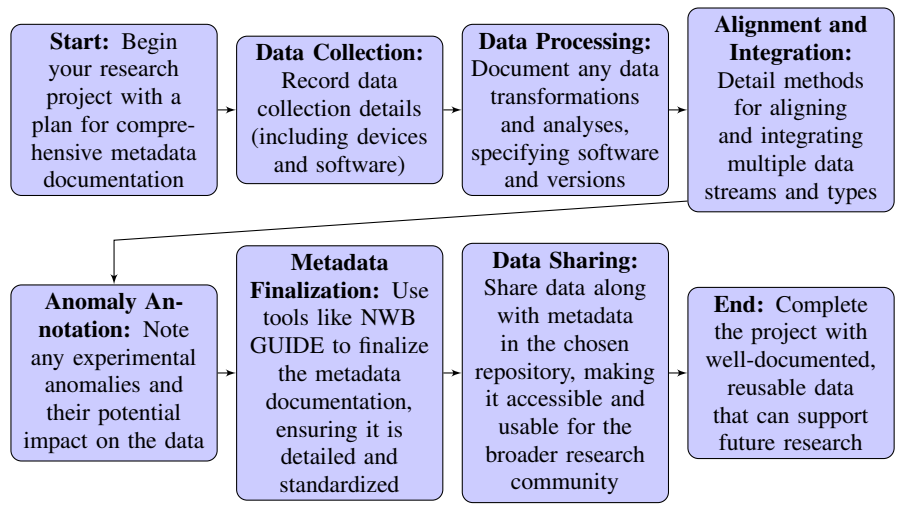

\paragraph{\textbf{Documentation and Metadata.}} 
\begin{itemize}
    \item Provide \textbf{thorough and structured metadata} to enable effective reuse of your data and use by researchers who are not familiar with your project. As AI and machine learning methods become increasingly integrated into neurophysiology data analysis, this will also ensure your datasets are \textbf{AI-ready} with rich metadata, greatly enhancing their reusability and the reliability of subsequent findings. See Fig. \ref{fig:docFlow} for steps of a recommended workflow when storing and sharing data. Specific recommendations about the types of data to include are expanded on below:
    \begin{enumerate}
        \item Document the \textbf{source script} and any other processes used to generate the dataset, even if they are not mandatory fields in your chosen data standard.
        \item Include comprehensive details about the \textbf{devices}, \textbf{software versions}, and \textbf{analysis algorithms} used during the experiments.
        \item Record any \textbf{stimuli} presented during the experiments, and include a detailed table specifying which stimuli were presented when.
        \item Clearly describe how \textbf{neural, behavioral and stimulus data streams} were aligned temporally.
        \item  Record key subject descriptors, like \textbf{genotype}, referencing external databases for standard definitions where applicable.
        \item Annotate any \textbf{anomalies or unusual occurrences} during data collection that might affect subsequent analyses.
    \end{enumerate}
    \item Utilize tools like \textbf{NWB GUIDE} for user-friendly and automated capture of important metadata, minimizing effort and enhancing standardization.
\end{itemize}

\paragraph{\textbf{Contributing to Existing Datasets.}} 

When running robust and well-validated analyses on existing datasets, consider sharing a derivative dataset comprising the results of these analyses. Contributing, for example, the results of dimensionality reduction analyses applied to a high-dimensional neural dataset or of pose estimation applied to behavioral recordings can greatly enhance the effective value of an existing dataset. For example, if the data is already shared on DANDI, addition of derivative to the existent dandiset or publishing it as a stand-alone dandiset is fairly simple and straightfoward.

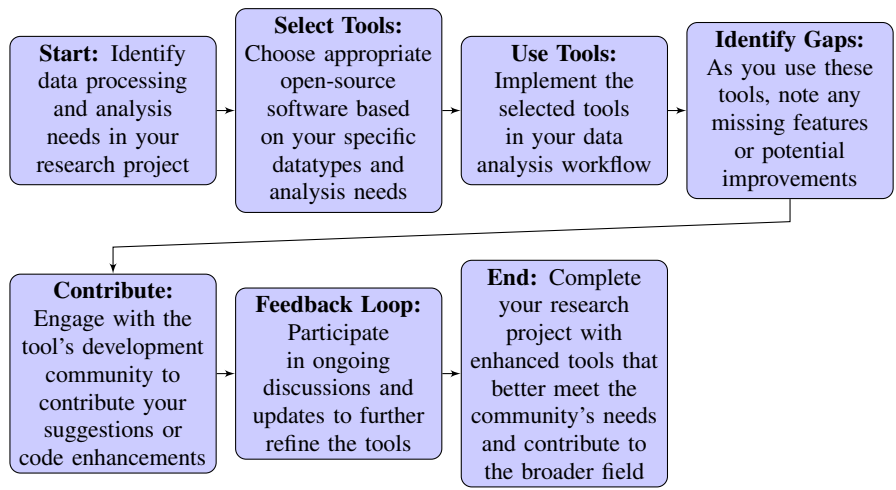
\begin{figure*}[hbt]
\centering
\begin{tikzpicture}[node distance = 3.5cm, auto]
    \tikzstyle{block} = [rectangle, draw, fill=myblue!30!white, 
    text width=8em, text centered, rounded corners, minimum height=4em]
    \tikzstyle{line} = [draw, -latex']
    \node [block] (start) {\textbf{Start:} Identify data processing and analysis needs in your research project.};
    \node [block, right of=start] (select) {\textbf{Select Tools:} Choose appropriate open-source software based on your specific datatypes and analysis needs.};
    \node [block, right of=select] (use) {\textbf{Use Tools:} Implement the selected tools in your data analysis workflow.};
    \node [block, right of=use] (identify) {\textbf{Identify Gaps:} As you use these tools, note any missing features or potential improvements.};
    \node [block, below of=start, yshift=-0.5cm] (contribute) {\textbf{Contribute:} Engage with the tool's development community to contribute your suggestions or code enhancements.};
    \node [block, right of=contribute] (feedback) {\textbf{Feedback Loop:} Participate in ongoing discussions and updates to further refine the tools.};
    \node [block, right of=feedback] (end) {\textbf{End:} Complete your research project with enhanced tools that better meet the community’s needs and contribute to the broader field.};
    \path [line] (start) -- (select);
    \path [line] (select) -- (use);
    \path [line] (use) -- (identify);
    \path [line] (identify.south) -- ($(identify.south) - (0,0.3)$) -- ($(start.south)!0.7!(contribute.north)$) -- (contribute.north);
    \path [line] (contribute) -- (feedback);
    \path [line] (feedback) -- (end);
\end{tikzpicture}
\caption{Tooling flowchart.}
\label{fig:toolFlow}
\end{figure*}

\paragraph{\textbf{Utilizing Existing Tools.}} Several steps can help optimize your use of existing tools. See Fig. \ref{fig:toolFlow} for an example workflow. Additional recommendations are provided below:
\begin{itemize}
    \item \textbf{Choosing Open-Source Software:} Due to the complexity of neurophysiological data analysis, it is advisable to use established open-source software packages, where possible and appropriate. These are less prone to errors and are continually vetted by the community. Examples include:
    \begin{enumerate}
        \item \textbf{Spike Sorting and Processing:} Consider tools like SpikeInterface and KiloSort.
        \item \textbf{Calcium Imaging Data Processing:} Consider tools like suite2p and CaImAn.
        \item \textbf{Pose Estimation:} Consider tools such as DeepLabCut and SLEAP.
    \end{enumerate}

    \item \textbf{Contributing to Tool Development:} If existing tools lack certain features or could be improved, contribute your enhancements back to the project. This type of collaboration:
    \begin{itemize}
        \item Allows the community to verify the robustness of the new feature.
        \item Enhances tool functionality and utility for the entire community.
        \item Accelerates scientific discovery and increases the robustness of research outcomes.
        \item Builds a culture of reuse and improvement, aligning with open science principles.
    \end{itemize}

\end{itemize}

\paragraph{\textbf{Developing New Tools.}} If you develop new tools from scratch, it can be very valuable to share these with the broader community. To maximize the robustness, usability, and findability of these tools, it is particularly helpful to:
\begin{itemize}
    \item \textbf{Share} the code on a platform like GitHub that enables robust version-control, as well as user feedback and contributions, ideally under a license that is highly permissive for code reuse and adaptation. 
    \item \textbf{Document} the code by including at minimum a README explaining the tool's intended use, the programming language it is designed in, its dependencies, usage examples, and ideally an interactive tutorial users can run in the cloud.  
    More detailed recommendations can be found in previously published articles \citep{Eglen2017}.
    \item Make a \textbf{plan} for long-term maintenance and promotion of the tool. This may require investing financial resources and hiring of dedicated personnel, but is generally critical for the longevity and usability of an open-source tool.
\end{itemize}

\subsection{Community-Wide Recommendations}


For major advances in open science practices to take hold in the field of neurophysiology, broad structural changes are needed that will improve the support and incentives offered to individual practicing neuroscientists. These changes will need to be implemented and adopted by the structures that oversee funding and resource allocation (i.e., funding agencies, funding committees, and universities), publication (i.e., journals, conferences, and their editorial boards), and academic career advancement (i.e., universities, departments, and tenure committees).

\paragraph{\textbf{Allocation of Funding and Resources.}} The adoption and implementation of open science practices can be very time consuming. It is therefore critical that laboratories have access to funding that is available or earmarked for hiring staff or students whose primary task is setting up and managing the laboratory's open science pipelines, tools, and datasets. It is also critical to ensure that laboratories have access to enough funding to maintain the equipment and tools required to create, deploy, and maintain their open science pipelines, tools, and datasets. 

At a broader scale, since many different laboratories in a same department, university, or region require similar open science resources, opportunities must be created for groups to secure stable and ongoing funding for the creation and maintenance of shared resources, also known as core facilities. Such resources could include long-term data storage facilities, high-performance computing clusters, and in-house research software engineering, data management, and data analysis support services. Access to internal or external expert consultants is a particularly under-considered need. Indeed, as discussed in an article of \textit{The Transmitter} \citep{Voigts2023}, the breadth and depth of skill required to run high-quality neuroscience studies has grown considerably over the last decades, to a level that is likely unreachable for most individual neurophysiology laboratories. To empower research laboratories to maintain a high quality of research, it is critical that researchers come together to identify the resources they need access to in order to best deploy their skills and allocate their time. Funding opportunities must therefore be readily available to address these resource needs. The creation of such shared resources would not only greatly alleviate pressure, reducing the breadth and depth of technical expertise required of individual laboratories, but could also greatly reduce resource waste by improving research quality across the board. 

\paragraph{\textbf{Recognition and Enforcement in Publications.}} It is important that we ensure that the value of open science contributions such as new datasets and tools is properly recognized in journals and conferences. To this end, clear submission and evaluation criteria must be developed enabling reviewers to fairly evaluate open science contributions for key features such as expected short-term and long-term usability and usefulness to the community. Developing effective criteria and directing reviewers to attend to these will not only improve recognition of these contributions to the field, but also draw wider attention to them.

Alongside improved recognition, mandatory adoption of open science practices must be more broadly enforced. In the long-term, a clear goal for the field is that all papers be published with their accompanying datasets and code made as public as ethics and privacy considerations allow. These datasets and codebases should be made available in formats that allow them to be easily deployed to confirm and extend the published results. Of course, these requirements cannot be enforced overnight as laboratories must be given the time and resources to build up their ability to meet them. Journals should nonetheless continue to progressively increase their adoption and enforcement of such policies. Early enforcement measures include directing reviewers to formally consider data and code sharing as part of their assessment of the submission quality. Actual data and code review could also begin to be included in the review process. To avoid overburdening reviewers, however, this would require authors to make their data and code very easy to access and deploy. Eventually, as open science practices become much easier to adopt, authors should be allowed to withhold data or code only when absolutely necessary, for example for ethical and privacy reasons.

Enforcement of open science practices will also need to come from the funding side of research. Major funding agencies have already taken measures requiring researchers to include a feasible data management plan when applying for funding \citep{NIH2025, Canada2025, ERC2025}. With these policies being comparatively new, however, it is not clear how carefully these plans are reviewed for feasibility and compliance, and to what extent and in what manner data sharing is enforced. As data sharing becomes the norm in the field, it will be important to establish clear and streamlined criteria for evaluating data management plans and for monitoring compliance with these plans. It may also be necessary to adopt penalties, like funding restrictions, for failure to comply with a data sharing plan without proper justification. Lastly, funding agencies should strongly consider extending these policies to code and software which are arguably often easier to share, and critical to fully understanding and assessing research results.

\paragraph{\textbf{Recognition in Research Career Evaluations.}} Finally, contributions to open science by members of the academic community, including undergraduate students, graduate students, postdoctoral fellows and professors, should be officially recognized as substantive contributions. Such contributions include, but are not limited to, standardizing and sharing datasets, and sharing well-organized and documented codebases, sharing and maintaining software tools, The impact of these shared resources can in turn be measured, for example, through citation counts, rates of reuse by the community, and the resource's role in advancing the field. Current efforts by the NIH to develop, through a community challenge, a data sharing index (the S-Index) represent an encouraging step in this direction \citep{Freelancer2024}, and will hopefully be accompanied by better recognition of other important open science contributions like software and codebase sharing. Data and code publishing platforms should also enable users to mint unique identifiers like DOIs (Digital Object Identifiers), streamlining the process by which contributions are cited and disseminated. Members of the academic community should also be encouraged to highlight their open science work in their CVs, university applications, and funding applications. Admissions committees, departments, hiring and tenure committees, and funding and grant review committees should in turn be directed to give these contributions proper credit in the evaluation of candidates' academic records.

\section{Concluding Remarks}

\begin{figure*}
\begin{tcolorbox}[colback=myblue!5!white, colframe=myblue, title=Key Takeaways for Advancing Open Science in Neurophysiology]
\begin{itemize}
\item \textbf{Education and training} of neuroscientists at every level is crucial for ensuring open data practices are effectively adopted and utilized.

\item \textbf{Funding investment} in the development, dissemination, and maintenance of open-source tools and infrastructure is necessary to support long-term sustainability and reliability of research outputs. Sponsors that value open data must be prepared to fund the health of the ecosystem, which includes supporting practicing neuroscientists, tool disseminators, and continuous maintenance/development to keep tools up to date.

\item \textbf{Improving research methodologies} by establishing benchmarks and standardized methodologies for data analysis and model evaluation will improve the reproducibility and comparability of research findings. Appropriately harnessing large language models (LLMs) and AI tools to enhance data analysis, literature search, and hypothesis generation could also significantly improve research quality and pertinence.

\item \textbf{Development of robust and user-friendly tools} for data management, analysis, and sharing is essential to support the adoption of open science practices across laboratories with varying resources.
Enhancing metadata quality and standardization is critical for the reusability and reproducibility of shared datasets, and comprehensive searchable metadata will greatly improving data utility. 
Such practices have the potential to help address important ethical considerations, like animal use in neurophysiology research, as optimal reuse of existing datasets can help keep animal use to a minimum.

\item \textbf{Alternative career paths} should be established within academia to support individuals skilled in data management and analysis, whose work is less focused on specific research hypotheses. Such positions would provide much needed job opportunities and security, while helping bridge the gap between traditional academic hierarchies and the increasingly complex technical landscape of neuroscience research.

\item \textbf{Progressive changes in culture and social infrastructure} are necessary and must occur alongside changes in incentives and credit assignment. Tool development and dataset contributions should receive greater appreciation and formal recognition (e.g. by hiring and promotion committees). Participating in this transformation is essential for fostering an environment that values and rewards open science.
\end{itemize}
\end{tcolorbox}
\caption{Key takeaways.}
\label{fig:takeaways}
\end{figure*}

This first ODIN symposium highlighted a growing momentum in neurophysiology research to incorporate the values and practices of open science.
The key takeaways from the symposium are summarized in Fig.~\ref{fig:takeaways}.
Innovations revolutionizing the quality and quantity of data we collect have been complemented by the development of robust data standardization and sharing platforms, along with a variety of computational resources for data processing, analysis and visualization. Large-scale data collection efforts are tackling the challenges of reproducibility and reliability in the field, with centralized approaches providing access to high quality data collection pipelines and decentralized ones encouraging collaborative protocol and analysis designs. However, significant challenges remain, particularly for laboratories with limited resources, where incorporating open science practices can be daunting and time-consuming. Furthermore, when laboratories do invest in adopting these practices, the time and effort required are often not sufficiently recognized by the traditional incentive structures of academic research.

Funding sponsors, publishers, and institutions wield the power to drive collaborative progress and sustain momentum. Their active support and recognition of the time and effort invested by researchers in open science initiatives are crucial for enabling this pivotal change in modern neuroscience. By acknowledging the value of open science practices, they elevate the entire field. Through incentivizing open data practices, funding robust infrastructure, and promoting tool dissemination, they create an environment where open science becomes a central pillar of neurophysiology research.

Meanwhile, we encourage researchers to actively engage in open science practices and leverage existing resources. By participating in the communities that build and use advanced tools, individuals can discover solutions to challenges they face, and tap into valuable community support. Where solutions are lacking, researchers can provide feedback reflecting their specific research needs, increasing the likelihood that future iterations will address those needs. Thus, while transformative impact arises from collective action with the much needed support of funders and institutions, it is essential to recognize the power of individual voices in shaping this action.

Overall, we anticipate gradual, collaborative progress in the field, rather than an overnight transformation, engaging researchers, sponsors and institutions. We advocate for acknowledging and celebrating symbiotic developments, which together will propel us toward more open, transparent, and impactful science. In this context, the ODIN symposium (intended as a bi-annual event) can serve as a vital platform for sustaining momentum, sharing novel developments, and addressing the evolving needs of the community.

\clearpage
\appendix*

\section{ODIN 2023 Talk Summaries}

\subsection{Devices, Neuroinformatics, and Platforms}

\subsubsection{New Devices and High Throughput Acquisitions}

\paragraph*{\textbf{Mapping the Human Brain with High Spatiotemporal Resolution.}}
Shadi Dayeh (University of California San Diego) detailed advancements in recording human brain activity using multi-thousand channel electrocorticography (ECoG) grids. He described the evolution from traditional clinical electrodes to modern microelectrode technologies that enable dense, high-resolution mapping. This advancement, facilitated by progress in thin-film microfabrication, allows for comprehensive brain activity mapping. Dayeh detailed the technological challenges that needed to be overcome to achieve this, such as scaling down electrode size to increase signal-to-noise ratio and adapting electrode design to enable stable contact with the brain's curvilinear surface \citep{Tchoe2022}.  He presented innovations like platinum nanorod grids (PtNRGrids) and highlighted the move towards wireless systems for efficient and less intrusive monitoring in both acute and chronic settings. \\

\paragraph*{\textbf{High Channel Count Electrophysiology: Present, future.}}
Neuropixels, a silicon probe which allows high-density simultaneous recording of hundreds of neurons in awake and freely moving animals, have revolutionized systems neuroscience \citep{Jun2017}. Tim Harris (HHMI Janelia Research Campus/JHU) discussed the development and applications of the latest version of the probe, Neuropixels NXT. Its more compact design further increases the detail and scale at which electrophysiological neural activity can now be recorded. Harris also discussed the challenges of handling the deluge of data Neuropixels probes produce, as well as data processing issues related to the fidelity of spike sorting. He noted the need for efficient data compression and management strategies to preserve the benefits of large-scale data collection. \\

\paragraph*{\textbf{Towards Cortex-Wide Recording of Neuroactivity at Cellular Resolution.}}
Alipasha Vaziri (Rockefeller University) showcased breakthroughs in developing technologies for cortex-wide, volumetric recording of activity in millions of neurons at single-cell resolution. Using techniques such as light sculpting and temporal multiplexing, his laboratory has been able to simultaneously record from millions of neurons. This type of data is poised to enable new insights into the complex dynamics of neuronal population activity and reveal patterns of coordinated activity across distant regions in the brain \citep{Demas2021}. Vaziri noted the challenges of running meaningfully interpretable analyses on such large datasets. However, he also shared evidence that the dimensionality of brain activity increases with the number of recorded neurons, indicating that the ability to record and meaningfully analyze neural activity across very large populations is indeed critical to understanding brain function. \\

\paragraph*{\textbf{Voltage Imaging: All-optical electrophysiology of neuron excitability.}}
Adam Cohen (Harvard University) presented an all-optical approach to electrophysiology in which neuron excitability is monitored through voltage imaging. By combining voltage-sensitive fluorescent proteins activated by red light with blue-light-activated channelrhodopsins for neuronal stimulation, his team has developed a neuro-optical interface for detailed monitoring of electrical activity, including spikes and subthreshold voltages, at high spatial and temporal resolutions \citep{Adam2019}. Cohen's work aims to dissect the complex dynamics that evolve within neurons and local networks by precisely measuring the input-output relationships and plasticity rules that govern changes in neural function. His talk underscored the difficulty of extracting clean signals from noisy data and of managing and sharing the voluminous datasets produced by these methods in the context of NIH data sharing mandates.

\subsubsection{Neuroinformatic Resources}
\paragraph*{\textbf{The Neurodata Without Borders Ecosystem for Neurophysiology Data Standardization: Driving collaboration in neuroscience.}}
Oliver R{\"u}bel (Lawrence Berkeley National Lab) described the role of Neurodata Without Borders (NWB)\footnote{For more information on NWB, see \url{https://www.nwb.org/}.} as a comprehensive data standard for the neurophysiology community. The standard was developed as a collaborative, multidisciplinary project under the NIH BRAIN Initiative with additional support provided by the Kavli foundation. R{\"u}bel detailed how NWB aims to address a wide range of neurophysiology data management needs by enabling diverse datatypes, from neural activity recordings to experimental metadata, to be organized, aligned, and integrated into a single, hierarchical, accessible format. He also discussed the growing NWB ecosystem, which now includes various tools and APIs for data conversion, inspection, and analysis, and emphasized ongoing enhancements such as support for cloud-based data access and integration with external resources \citep{Rubel2022}. \\

\paragraph*{\textbf{DANDI: An archive and collaboration space for neurophysiology projects.}}
Satrajit Ghosh (Massachusetts Institute of Technology (MIT)) presented the data repository DANDI (Distributed Archives for Neurophysiology Data Integration)\footnote{For more information on DANDI, see \url{https://dandiarchive.org/}.}. DANDI is supported by the Brain Initiative and Amazon Web Service (AWS) public dataset program, and operated in collaboration with MIT, Kitware, and Catalyst Neuro. Its core aim is to make neurophysiology data, including but not limited to electrophysiology and optophysiology data, readily accessible and usable for the research community. Ghosh emphasized that DANDI, with its cloud-based infrastructure, is committed to hosting the largest collection of neurophysiology data globally. In alignment with FAIR principles, data hosted on DANDI must be shared in standardized formats, such as NWB and BIDS, under a Creative Commons license. Ghosh also discussed the ongoing development of tools to facilitate data submission, collaboration, and analysis. He underscored the need for computing resources, shared ecosystems, and trainings on data sharing to be broadly available to the community. He noted that with the breadth of datatypes it hosts, DANDI provides a collaborative space for bringing together neurophysiology data from diverse research areas. As such, the vision for DANDI is to extend beyond data storage to foster a comprehensive ecosystem for neuroscience research. \\

\paragraph*{\textbf{End-to-end Computational Workflows for Neuroscience Research.}}
Dimitri Yatsenko (DataJoint)\footnote{For more information on DataJoint, see \url{https://github.com/datajoint}.} focused on the development and application of end-to-end computational workflows in neuroscience research \citep{Yatsenko2021}. He outlined the importance of considering an entire project lifecycle-from data acquisition to analysis-and described diverse needs that arise in neuroscience research, from animal management and behavior monitoring to electrophysiology, spike sorting, and behavior analysis. Yatsenko presented the concept of ``operational maturity'' \citep{Johnson2023} in neuroscience research to illustrate how DataJoint Elements's open-source solutions can increase the efficiency and reproducibility of a laboratory’s workflows. In addition to aligning with FAIR principles, the adoption of standardized workflows can improve collaboration within the neuroscience community. Examples like the creation of an interactive environment for working with the MICrONS dataset \citep{Microns2023}, and the coordination and automation of various collaborative projects illustrate the practical application of workflow tools in elevating the operational capabilities of neuroscience laboratories. \\

\paragraph*{\textbf{Web-based Visualization and Analysis of Neurophysiology Data.}}
Jeremy Magland (Flatiron Institute) introduced open-source software tools for web-based visualization and analysis of neurophysiology data. He presented three main tools he has been developing: \textit{Figurl}\footnote{For more information on Figurl, see \url{https://github.com/flatironinstitute/figurl}.}, a framework for creating and sharing interactive visualizations, \textit{Neurosift}, a tool for browsing NWB files, particularly those hosted on DANDI \citep{Magland2024}, and \textit{Dendro}\footnote{For more information on Dendro, see \url{https://dendro.vercel.app/}.}, a prototype web app for analyzing neurophysiology data in the cloud, or using local or cluster compute resources. Magland detailed how these tools can facilitate scientific collaboration, reproducibility, and knowledge transfer by, for example, simplifying the sharing of interactive figures and data visualizations through URLs generated by Python scripts. The value of integrating of these tools with data standards like NWB and platforms like DANDI was emphasized, as were the benefits of client-only applications which do not require server maintenance.

\subsubsection{Platforms \& Collaborative Initiatives}

\paragraph*{\textbf{Brain Mapping and Disease Modellings using Genetically Modified Marmosets.}}
Hideyuki Okano (Keio University/Riken) presented Japan's Brain/MINDS project for brain mapping and disease modeling using genetically modified marmosets. The initiative has generated a comprehensive, publicly shared dataset that includes structural, diffusion and resting-state MRI datasets, as well as quantitative 3D data, and an \textit{in situ} hybridization-based marmoset gene atlas \citep{Hata2023,Okano2015,Skibbe2023}. By integrating gene expression and brain structure data, this comprehensive database serves as a valuable reference for detecting abnormalities in disease models and facilitating interspecies comparisons. The initiative's research has notably revealed, among other discoveries, between and within-column connectivity patterns in the prefrontal cortex of marmosets that are not observed in mice. The initiative has also developed models of certain neurodegenerative and neurodevelopmental diseases. This has enabled, for example, a detailed study of Rett syndrome using CRISPR Cas9, which revealed reduced connectivity, poor dendritic arborization, and a disruption in excitatory/inhibitory balance due to hypermaturation of parvalbumin neurons. These studies also showed that MECP2 knock-out marmosets display gene expression changes similar to those observed in human patients, bolstering the potential of this research to provide insights into the molecular mechanisms underlying Rett syndrome, and to reveal potential therapeutic targets. \\

\paragraph*{\textbf{OpenScope: The first astronomical observatory in neuroscience.}}
 J{\'e}r{\^o}me Lecoq (Allen Institute for Neural Dynamics) introduced the Allen Brain Observatory, a vast database of cellular-level activity in the mouse visual system, and its OpenScope platform, which serves as a shared resource for generating high throughput and reproducible neurophysiology data. Lecoq described how these initiatives have evolved as pipelines developed for two-photon microscopy and Neuropixels recordings into comprehensive platforms to which external scientists can submit project proposals for data collection. The operational model, which uses a double-blinded review process, seeks to ensure equitable and merit-based access akin to how time is allocated on astronomical telescopes \citep{DeVries2023,Koch2022}. In so doing, OpenScope aims to be highly equitable and inclusive of researchers who would not otherwise have access to such resources. Lecoq also shared the development of the OpenScope Databook, which aims to democratize and help standardize data visualization and analysis tools across the community. He further emphasized the substantial financial investment required to run these high-caliber projects. \\
 
\paragraph*{\textbf{Compute, Data \& Standards in Large-Scale Neuroscience.}}
David Feng (Allen Institute for Neural Dynamics) discussed how computing, data management, and standards are approached within the context of large-scale neuroscience research at the Allen Institute for Neural Dynamics. Feng outlined the Institute's ambitious mission to uncover the neural underpinnings of emotions, memories, and actions, utilizing advanced neurotechnology tools to simultaneously comprehensively capture brain-wide activity and extensive behavioral data. Emphasizing a commitment to open science, he described efforts to ensure data are immediately FAIR by ensuring robust, machine- and human-readable metadata are captured at the time of acquisition. He highlighted how cloud computing enhances the utility and inclusiveness of open science initiatives, reducing the logistical challenges associated with moving and storing terabytes to petabytes of data. Cloud development and sharing platforms like \textit{``GitHub Codespaces''} and \textit{``Code Ocean''} also simplify the sharing of complete software and hardware environments, making reproducible science feasible on a large scale. In this vein, Feng also discussed the usefulness of containerized NextFlow pipelines which are designed to enable fully reproducible data processing across cloud and on-premise environments. \\

\paragraph*{\textbf{International Brain Laboratory: A brain-wide map of neuronal activity during behavior.}}
Matteo Carandini (University College London) presented the International Brain Laboratory (IBL), a collaborative effort involving 22 laboratories across many countries to create a brain-wide map of neuronal activity during behavior in mice. Carandini explained how the IBL has standardized experimental protocols to overcome reproducibility challenges and pooled data from nearly 33,000 neurons using Neuropixels probes. The resulting dataset enables a comprehensive analysis of how different brain regions contribute to sensation, decision-making, and actions, and encode prior beliefs \citep{IBL2021,IBL2023,Findling2023}. Carandini also touched on challenges encountered when analyzing IBL data. In particular, despite achieving significant reproducibility and uncovering consistent and widespread patterns in how behavioral information is encoded across the brain, the team found that different analytical methods used on the same data could nevertheless yield distinct results and interpretations \citep{IBL2024}\footnote{For more information on the IBL, see \url{https://www.internationalbrainlab.com/}.}.

\subsection{Knowledge Extraction, Software, Modeling}

\subsubsection{Keynote: Linking Large-scale Neural Data to Behavior: Algorithms \& Opportunities.}
Mackenzie Mathis (École Polytechnique Fédérale de Lausanne) emphasized the importance of approaching neurophysiology and animal behavior data through a shared scientific lens. She demonstrated the pivotal role cutting-edge computational algorithms and AI play in enabling researchers to decipher the relationship between neural activity and complex behaviors. Mathis presented DeepLabCut \citep{Mathis2018}, an AI toolkit that automates pose estimation from animal recordings\footnote{For more information on DeepLabCut, see \url{https://github.com/DeepLabCut}.}, and introduced CEBRA \citep{Schneider2023}, a tool for joint embedding of behavioral and neural data\footnote{For more information on Cebra, see \url{https://cebra.ai/}.}. Her talk underscored key ways to support tool longevity and broader community adoption, in order to democratize advanced computational analysis. \\

\subsubsection{Experimental Models and Data-Driven Approaches}
\paragraph*{\textbf{Intrinsic Activity In Human Cortical Organoids Reveal Protosequences that Model Default States in the Developing Cortex.}}
Kenneth Kosik (University of California, Santa Barbara) described how human cortical organoids can be used to study network development and electrical signaling within neural circuits, specifically through the use of integrated optofluidic-CMOS multielectrode arrays \citep{Sharf2022}. He argued that organoids have a great potential to enable new in vitro discoveries as they display properties, like lamination and local field potentials, observed in the brain, but not in traditional two-dimensional neuronal cultures. Kosik detailed the use of technologies like high-density MEA arrays (MaxWell Biosystems) and Neuropixels to study the intrinsic activity of the cortex in the absence of sensory input, and he explored the potential of using repeated stimulation to mimic learning processes in organoids. \\

\paragraph*{\textbf{Data-driven Dynamic Models of Large-scale Neural Data.}}
Bing Brunton (University of Washington) demonstrated how data-driven models can be used to decode and understand large-scale neural recordings. This work employs video-annotated human electrocorticography recordings, collected in a natural setting, to investigate the relationship between natural behaviors and brain activity, with implications for improving brain–computer interface technologies \citep{Peterson2021, Peterson2022}. Public access to this dataset, called AJILE12, has exemplified the transformative power of open science, as it has enabled researchers and students worldwide to access and analyze real data\footnote{For more information on AJILE12, see \url{https://github.com/neurovium/Neuromatch-AJILE12}.}. \\

\paragraph*{\textbf{A Less Artificial Intelligence: Exploring mechanisms through MICrONS.}}
Andreas Tolias (Stanford University) presented the MICrONS project, a collaborative scientific endeavor which provides an open and publicly accessible data portal for accessing connectivity and functional imaging data collected by a consortium of laboratories. These data include large-scale electron microscopy-based reconstructions of cortical circuitry from mouse visual cortex, along with corresponding functional imaging data from some of those same neurons\footnote{For more information on MICrONS, see \url{https://www.microns-explorer.org/}.}. Tolias also discussed the important question of how the brain is able to generalize from limited data so much more successfully than current AI systems. He described how ``inception loops'' where neural networks are trained on neurophysiological recordings to reveal the stimuli that maximally excite specific neurons \citep{Walker2019}, can be used to better understand complex systems like the visual processing system. \\

\paragraph*{\textbf{The Role of Inhibitory Neurons in Auditory Processing.}}
Maria Geffen (University of Pennsylvania) discussed the influence of inhibitory neurons on auditory processing and perception. Her laboratory uses opto-electric recordings to analyze how different types of inhibitory neurons modulate frequency discrimination and adaptation to temporal patterns in complex sound processing \citep{Natan2015, Tobin2025}. Geffen’s work highlights the fundamental role of inhibitory neurons in discerning frequencies and in shaping the auditory cortex’s response to complex sound patterns.

\subsubsection{Neuroscience Toolkits}
\paragraph*{\textbf{MoSeq (Motion Sequencing): Quantifying 3D video of freely behaving animals.}}
Bob Datta (Harvard Medical School) introduced MoSeq (Motion Sequencing), a tool for parsing 3D videos of freely behaving animals in naturalistic settings. MoSeq utilizes depth cameras to capture detailed 3D movements of rodents, identifies distinct behavioral ``syllables'', and constructs behavioral state maps offering insights into the sequential and contextual structure underpinning natural behaviors \citep{Wiltschko2020}. This unsupervised machine learning approach allows users to dissect the intricate patterns underlying animal behavior, and to explore the impact of various external perturbations on these behavioral patterns. Datta highlighted MoSeq's potential contributions to neuroscience research, amongst other things through its ability to distinguish between the effects of different drugs on behavior and reveal variability in behavior across individuals that is both significant and consistent over time. He emphasized that MoSeq is open and accessible, as well as robust and versatile, making it a significant contribution to the field, poised to gain wide adoption and continue to develop through collaborative contributions\footnote{For more information on MoSeq, see \url{https://dattalab.github.io/moseq2-website}.}. \\

\paragraph*{\textbf{SpikeInterface: Spike sorting in large-scale recordings.}}
Alessio Buccino (Allen Institute for Neural Dynamics) presented SpikeInterface, a collaborative Python package designed to simplify and standardize the spike sorting step in electrophysiology data processing \citep{Buccino2020}. Buccino detailed the challenges faced in the field, such as the wide variety of acquisition systems and file formats, along with the lack of reproducibility across toolboxes and the frequent lack of data provenance information. SpikeInterface addresses these by providing a unified interface for comparing the outputs of various spike sorting algorithms and pre-processing tools, applying them to data and generating detailed pre-processing reports. SpikeInterface supports over 15 spike sorters and facilitates the entire spike sorting pipeline, from pre-processing and sorting to post-processing and visualization, all while being open to community development\footnote{For more information on SpikeInterface, see \url{https://github.com/SpikeInterface}.}. \\

\paragraph*{\textbf{Machine Learning Tools for Understanding Complex Hippocampal Patterns in Learning and Memory.}}
Andrea Navas-Olive (IST-Austria) presented new machine learning tools for analyzing electrophysiological data, with a specific focus on sharp wave-ripples (SWRs), which are crucially involved in memory consolidation. She discussed challenges in detecting SWRs due to the variability in their statistics, and how traditional spectral methods might bias the types of events that are detected. Employing machine learning techniques, she and her colleagues have developed algorithms that improve detection, while reducing dependency on often arbitrarily selected thresholds which can bias the characteristics of detected events \citep{Navas-Olive2022}\footnote{For more information on SWR detection tools, see \url{https://github.com/PridaLab/cnn-ripple} and \url{https://github.com/PridaLab/rippl-AI}.}. These algorithms are not only applicable to different areas of the brain but also generalize to other species, demonstrating their potential for broad applications in neuroscience research \citep{Navas-Olive2024}. The innovative aspects of this work extend beyond the development of sophisticated detection algorithms: they also encompass the collaborative, crowd-sourced approach to problem-solving behind the project. BrainCode Games hackathon engaged a diverse group of participants from various backgrounds to tackle the challenge of SWR detection. This collaborative effort not only led to the creation of multiple effective machine learning models, but also fostered the development of a community of interdisciplinary researchers and industry professionals united by a common goal. The hackathon's success highlights the value of community-driven research endeavors for generating unbiased, comprehensive solutions to shared problems. \\

\paragraph*{\textbf{Neural Ensemble \& HBP/EBRAINS Knowledge Graph.}}
Andrew Davison (Université Paris-Saclay/Centre National de la Recherche Scientifique) reviewed the accomplishments of the Human Brain Project (HBP) and its integration in the EBRAINS Knowledge Graph, two initiatives aimed at advancing digital neuroscience. The HBP, a decade-long EU-funded project with a budget of approximately 600 million Euros brought together over 500 researchers and 100 universities to push the boundaries of science and engineering in pursuit of understanding the human brain. A culminating outcome of this project was the establishment of EBRAINS, a comprehensive research infrastructure aggregating digital tools and services for neuroscientists. Davison highlighted the transformative potential of EBRAINS for facilitating data-driven science, emphasizing its role in connecting the neuroscience community to sustainable, high-quality digital resources. The EBRAINS Knowledge Graph, a core component of this infrastructure, serves as a universal metadata repository for the project, enhancing resource discoverability and interoperability. This digital repository enables the sharing of experimental data, computational models, and software tools, all interconnected within a semantic, linked data framework \citep{Appukuttan2022,Bologna2022}. Through its sophisticated architecture and community-driven approach, the EBRAINS initiative exemplifies the importance of collaborative science and open data when tackling the most complex questions in neuroscience\footnote{For more information on EBRAINS, see \url{https://www.ebrains.eu/}.}.

\subsubsection{Modeling and Benchmarking}
\paragraph*{\textbf{Brain-Score (Integrative Benchmarks For Models at Scale).}}
Martin Schrimpf (École Polytechnique Fédérale de Lausanne) explained how the Brain-Score platform\footnote{For more information on Brain-Score, see \url{https://www.brain-score.org/}.} harnesses large-scale datasets to construct integrative benchmarks that evaluate neural and behavioral models. By linking disparate datasets through systematic benchmarking, Brain-Score provides unified constraints that guide the development of improved models and enable prediction of future experimental outcomes \citep{Tuckute2024,Schrimpf2024,Bashivan2019,Schrimpf2018,Schrimpf2020}. \\

\paragraph*{\textbf{Taming Machine Learning Models of Neural Dynamics with Anatomical and Behavioral Constraints.}}
Shreya Saxena (Yale University) emphasized how incorporating biophysical and anatomical constraints into neural network training enables models to better capture the adaptive dynamics of the motor cortex, and more broadly recapitulate the neural activity observed during movement and interactions between subjects. Such models can then provide insight into the neural encoding strategies that underlie flexible motor control, i.e., how the motor cortex adapts its activity patterns to drive different muscle movements as tasks change. Specifically, the representational changes observed can be studied to understand how they guide the activation patterns of different muscle groups that allow animals to make precise adjustments to their movements \citep{Saxena2022,Almani2022,Almani2024}. \\

\paragraph*{\textbf{Low-dimensional Manifolds for Neural Population Dynamics.}}
Hannah Wirtshafter (Northwestern University) explored the dynamics of learning and memory representations in the hippocampus, particularly focusing on how they evolve across spatial contexts. Using calcium imaging and open-source analysis tools in a Trace Eyeblink Conditioning task, which is hippocampal dependent, she observed that animals were able to rapidly reapply conditioned responses when transitioning between environments, despite extensive place cell remapping occurring in their hippocampus.  Manifold analyses suggested, however, that task encodings remained stable even following remapping \citep{Wirtshafter2024}. Wirtshafter's work notably compared a number of dimensionality reduction techniques, highlighting the importance of ensuring researchers have access to easily deployable analytical tools, as well as high quality documentation clarifying the differences, applicability, and limitations of different tools. \\

\paragraph*{\textbf{Quantifying Animal-to-animal Variability in Large-Scale Neural Recordings through Shape Metrics.}}
Alex Williams (Flatiron Institute/NYU) highlighted both the challenge and potential value of quantifying animal-to-animal variability in large-scale neural recordings for advancing systems neuroscience research. Leveraging large datasets from initiatives like the IBL, through which standardized data was collected across multiple laboratories, Williams demonstrated how ``shape metrics'' can be used to compare neural representations or manifolds across different animals in a high-dimensional space, irrespective of individual differences in neuron populations \citep{Williams2021,Duong2023, Pospisil2024, Harvey2023}. Drawing on principles from shape theory, he aims to develop open-source tools for analyzing neural data that go beyond traditional R-squared scores, and instead allow models to be matched to specific hypotheses about the neural basis of behavioral variability.

\subsection{Breakout Sessions: Beyond `FAIR': What does sustainable protocolization of open data in neuroscience look like?}
Joseph Monaco, Scientific Program Manager in the Office of the BRAIN Director of NIH BRAIN Initiative, opened the breakout sessions highlighting a vision for a sustainable, open data ecosystem that builds on the BRAIN 2025 core principles. Central to this vision is the creation of public, integrated repositories for datasets and analysis tools. The envisioned BRAIN Data Ecosystem would provide infrastructure for seamless data coordination, integration, interoperability, and reuse, promoting the core open science aims of data sharing, reproducibility, and rigor. Another aim is the gradual introduction of mandates, such as the BRAIN Data Sharing Policy, requiring data and code to be shared alongside publications\footnote{For more information on BRAIN 2025, see \url{https://braininitiative.nih.gov/vision/nih-brain-initiative-reports/brain-2025-scientific-vision}.}. This initiative aims to ensure that research data are made available to the wider community in a timely manner. The proposed framework spans a wide range of data domains—including light microscopy, multi-omics, neurophysiology, and human neuroimaging—and is distributed across platforms such as BossDB, NeMO, DANDI, OpenNeuro, and DABI (see Tables \ref{tab:brain_archives} and \ref{tab:archives}). Collectively, these repositories house thousands of datasets, demonstrating the vast scale of neuroscience data currently being shared and analyzed. With targeted funding opportunities and strategic mission goals, this sustainable data ecosystem is poised to foster advances in data science and deepen our understanding of the brain, ultimately paving the way for groundbreaking discoveries in brain health and disease.

%

\begin{acknowledgments}
The organizers of ODIN would like to thank all the speakers for their engaging presentations. We also thank all the participants for their thoughtful engagement in the ODIN symposium and ensuing discussions. We thank the discussion moderators, including Ben Dichter, Yaroslav Halchenko, Katherine Fairchild, Mark Harnett, Joseph Monaco, Matthew Wilson, Ila Fiete, Horea Christian, Dorota Jarecka, William Broderick, and Guillaume Viejo, for guiding these discussions. We wish to acknowledge support for the ODIN symposium from DANDI and MIT McGovern Institute for Brain Research. This work is supported by NIH Grant R24MH117295. C.J.G. is supported by the Wellcome Trust (Wellcome Trust Investigator Award 200790/Z/16/Z awarded to Claudia Clopath). R.L. is supported by NIH grant U24NS120057 and The Kavli Foundation. M.S. is supported by NIH grant U19NS132720, a C.V. Starr Fellowship, and a Burroughs Wellcome Fund's Career Award at the Scientific Interface.
\end{acknowledgments}

\bibliographystyle{apalike}
\bibliography{odin}

\end{document}